\documentclass[11pt]{article}

\usepackage{amsmath,amsfonts,amssymb,amsthm}
\usepackage{graphicx}
\usepackage{xcolor}
\usepackage{url}
\usepackage{geometry}
\usepackage{authblk}
\usepackage{float}
\usepackage{dcolumn}
\usepackage{multirow}
\usepackage{authblk}

\usepackage[skip=0.5ex]{subcaption}
\usepackage{hyperref}

\hypersetup{
    colorlinks=true,
    linkcolor=blue,
    citecolor=red,
}
\usepackage{cleveref}
\crefname{figure}{Figure}{Figs.}

\numberwithin{equation}{section}
\numberwithin{figure}{section}

\newcommand{\rom}[1]{\uppercase\expandafter{\romannumeral #1\relax}}
\newcommand{\figref}[1]{Fig.~\ref{#1}}
\begin{document}

\definecolor{TW-color}{RGB}{100,0,100}
\definecolor{MR-color}{RGB}{0,0,255}
\definecolor{SK-color}{RGB}{0,255,0}
\definecolor{Error-color}{RGB}{250,0,0}
\newcommand{\TWedit}[1]{{\color{TW-color}#1}}
\newcommand{\Error}[1]{{\color{Error-color}#1}}
\newcommand{\SKedit}[1]{{\color{SK-color}#1}}
\newcommand{\MRedit}[1]{{\color{MR-color}#1}}

\graphicspath{{./images/}}

\title{\textbf{Scattering of Vortices with Excited Normal Modes}}
\author[1]{Steffen Krusch\thanks{\tt S.Krusch@kent.ac.uk}}
\author[1]{Morgan Rees\thanks{\tt mjrr2@kent.ac.uk}}
\affil[1]{\it{\small{School of Mathematics, Statistics and Actuarial Sciences, University of Kent, Canterbury, United Kingdom}}}
\author[2]{Thomas Winyard\thanks{\tt twinyard@ed.ac.uk}}
\affil[2]{\it{\small{Maxwell Institute of Mathematical Sciences and School of Mathematics, University of Edinburgh, Edinburgh, United Kingdom}}}
\date{}

\allowdisplaybreaks

\maketitle
\begin{abstract}
We consider head-on collisions at critical coupling of vortices modelled by the Abelian-Higgs model. We investigate the 2-vortex scattering, whereby the vortices are excited by the shape mode causing fluctuations in the gauge-invariant quantities. When the vortices are excited with a sufficiently large amplitude the moduli space approximation fails, and we observe an interesting behaviour in which the vortices can become trapped in a quasi-bound state with multiple bounces. We perform a detailed investigation on the behaviour of these excited vortices and sample a phase space of solutions. Interestingly, we find a fractal structure dependent on the initial phase of the mode and velocity of the vortices. 
\end{abstract}

\section{\label{sec:intro} Introduction}
The Abelian-Higgs model \cite{higgs1966spontaneous} is a relativistic field theory whose excitations in $(2+1)$-dimensions take the form of topologically stable solitons known as vortices. The field theory consists of a complex scalar field $\Phi$ coupled to a $U(1)$ gauge field $A_\mu$. The static theory is equivalent to the effective Ginzburg-Landau theory \cite{ginzburg1950j}, describing a magnetic field penetrating a superconductor, quantised by the number of vortices. The dynamics of vortex solutions is where these two theories diverge; the Abelian-Higgs model exhibits $2^{\rm nd}$ order dynamics with Lorentz invariance \cite{moriarty1988dynamical,myers1992study,shellard1988vortex}, whereas the time dependent Ginzburg-Landau model exhibits $1^{\rm st}$ order dynamics \cite{dorsey1992vortex,manton1997first}. It is the former $2^{\rm nd}$ order dynamics that we will focus on in this paper. Note that in $(3+1)$ dimensions vortices appear as string-like objects, coined cosmic strings, that if they exist, may be detected through the gravitational contribution to early universe cosmology \cite{vilenkin1985cosmic}.

Vortex scattering has been well studied for all values of the single parameter $\lambda$  \cite{moriarty1988dynamical,myers1992study,shellard1988vortex,SteffenJenny,Steffen2}. This parameter splits the model into two types; type I ($\lambda < 1$) where vortices exhibit long-range attraction and type II ($\lambda > 1$) where vortices repel at long-range. In contrast, at critical coupling ($\lambda = 1$), there are no static long-range inter-vortex forces and the $N$-vortex solutions can be represented by an unordered set of dimension $2N$ or $\mathcal{M}_N = \mathbb{C}^N/S_N$ where $S_N$ is the set of permutations. At critical coupling, the low energy $2^{\rm nd}$ order dynamics can then be approximated as free geodesic motion on the moduli space $\mathcal{M}_N$. This moduli space naturally captures the most striking result, namely,  that vortices exhibit head on $90^{\circ}$  scattering \cite{taubes1}.

In this paper we will consider the $2^{\rm nd}$ order dynamics of vortices away from $\mathcal{M}_N$ by exciting normal modes of the individual vortices. We will demonstrate that vortices do exhibit long-range forces at critical coupling when their normal modes are excited. These long-range interactions alternate between attractive and repulsive depending on the phase of the excited normal mode. We will then consider the effect of these excited modes on the scattering of vortices. 

Several studies have considered the effect of excited normal modes on the scattering of solitons and anti-solitons in 1-dimension, coined wobbling kinks \cite{kinkWobble}. The scattering of wobbling kink/anti-kinks (while exhibiting strong attractive static forces) are shown to bounce off each other depending on initial velocities and the amplitudes of the excited mode. The number of bounces has also been shown to be chaotic in nature.

The paper is organised as follows. Section \ref{sec:model} outlines the Abelian-Higgs model. Section \ref{sec:normalModes} discusses the excitation of the shape mode. Section \ref{sec:numerics} explores the numerical techniques used in simulating vortex dynamics. Section \ref{sec:results} discusses the results we find from scattering vortices with excited shape modes. Finally, we conclude in Section \ref{sec:conclusions} with a summary of the results, as well as proposing some future ideas to be considered. 

\section{\label{sec:model} The model and static vortex solutions}
The $(2+1)$-dimensional Abelian-Higgs model \cite{higgs1966spontaneous} is described by the following Lagrangian,
\begin{equation}
L = \int_{\mathbb{R}^2}\left\{\frac{1}{2}\overline{D_{\mu}\Phi}D^{\mu}\Phi -\frac{1}{4}f_{\mu\nu}f^{\mu\nu} - \frac{\lambda}{8}\left(1-|\Phi|^2\right)^2\right\}\ d^2x\,,
 \label{eq:Lagrangian}
\end{equation}
where $\Phi(t,x)$ is a complex scalar field called the Higgs field. The $U(1)$ gauge potential is denoted by $A_{\mu}(t,x)$, which is a real one form where $A_{\mu} = (A_0, A_1, A_2),$ and $D_\mu = \partial_\mu - iA_\mu$ is the associated covariant derivative. Moreover, we denote the 2-form field strength tensor as $f_{\mu\nu} = \partial_{\mu}A_{\nu} - \partial_{\nu}A_{\mu}$, where $f_{12}$ gives the local magnetic field orthogonal to the plane. We will ensure that $\mathbb{R}^{2+1}$ has the signature $(+,-,-)$ throughout the paper. Finally, as a gauged theory, \eqref{eq:Lagrangian} is invariant under the following gauge transformation,
\begin{align}
\Phi(x) &\mapsto e^{i\alpha(x)}\Phi(x)\,,&
A_{\mu} &\mapsto A_{\mu} + \partial_{\mu}\alpha(x)\,.
\end{align}
Note that we have rescaled the model to normalise all parameters (e.g. the electric charge or vacuum expectation value of the Higgs field) leaving a single parameter, the mass of the Higgs field $\lambda.$ The speed of light is also set to $c = 1.$
The static energy can then be written as
\begin{equation}
    \label{eq:staticEnergy}
    V[\Phi,A_i] =\frac{1}{2}\int_{\mathbb{R}^2}\left\{\overline{D_i\Phi}D_i\Phi + B^2 + \frac{\lambda}{4}(1-|\Phi|^2)^2\right\}\ d^2x\,,
\end{equation}
where $B$ is the magnetic field, such that $B = f_{12} = \partial_1 A_2 - \partial_2 A_1$.

Varying the Lagrangian with respect to the independent fields $(\Phi,A)$, we obtain the resulting $2^{\mathrm{nd}}$ order dynamic non-linear equations of motion,
\begin{align}
D_{\mu}D^{\mu}\Phi - \frac{\lambda}{2}(1 - |\Phi|^2)\Phi &= 0\,,\label{eq:eom1}\\	
\partial_{\mu} f^{\mu\nu} + \frac{i}{2}(\overline{\Phi}D^{\nu}\Phi - \Phi\overline{D^{\nu}\Phi}) &= 0\,.
 \label{eq:eom2}
\end{align}
For field configurations to have finite energy we require that $B \to 0\,,$ $D_{\mu}\Phi \to 0$ and $|\Phi| \to 1$ as $\rho \to \infty\,,$ where $\rho = |x|$. This fixes the Higgs field on the boundary 
$$
\Phi_\infty := \lim\limits_{\rho\to\infty}\Phi(x)
$$ 
to take values on the unit circle such that $\Phi_\infty: S^1_\infty \to S^1$, where $S^1_\infty$ is the circle on the boundary of $\mathbb{R}^2$. This map is encapsulated by an integer degree or winding number $N\in\mathbb{Z}$. This winding number defines the number of zeros of the continuous Higgs field $\Phi$ including multiplicity. 
Since a given field configuration cannot be deformed from one homotopy class into a different one by a continuous deformation, the field configurations are separated into infinitely many disjoint components, indexed by the integer degree $N$. 
Therefore, the dynamic field equations must preserve the integer degree $N$. Using the boundary conditions above and Stokes' theorem,
we can write the total magnetic flux in terms of the degree $N$ which is hence quantised,
\begin{equation}
-\frac{1}{2\pi} \int_{\mathbb{R}^2} f_{12} = N.
 \label{TopCharge}
\end{equation}
We will interpret the degree $N$ as counting the number of vortices in the plane, the positions of which are taken to be the zeroes of the Higgs field $\Phi$.

In this paper we are interested in the critically coupled case ($\lambda = 1$) for which it can be shown that the static energy is bounded below by the degree $E[\Phi,A] \geq \pi |N|$, called the Bogomolny bound \cite{bogomol1976stability}. This inequality is saturated if and only if the fields satisfy the Bogomolny equations,
\begin{align}
\label{Bogeq}
D_1 \Phi + i D_2 \Phi &= 0\,,&
f_{12} + \frac{1}{2}\left( 1 - |\Phi|^2\right) & = 0\,.
\end{align}
This set of $1^{\rm st}$ order equations were studied in detail by Taubes \cite{taubes1,taubes2}. The moduli space $M_N$ is defined as the space of all possible vortex configurations of topological charge $N$ that satisfy the equations \eqref{Bogeq} and thereby minimise the energy functional. In fact, the moduli space is a manifold of dimension $2N,$ and the kinetic energy induces a natural metric on $M_N,$ \cite{Samols}. 
When vortices have small velocities, their scattering dynamics can be well approximated by geodesic motion on the moduli space $M_N,$ \cite{Stuart}. The moduli space metric needs to be evaluated numerically. However, there are analytic results for vortices in hyperbolic space, \cite{Strachan, Steffen1}.

The moduli space approximation correctly predicts that two vortices which approach each other head-on will scatter at right angles which we will confirm numerically in the following section.

We first consider an axially symmetric static isolated vortex of degree $N$ at the origin using the following ansatz,
\begin{align}
\label{eq:polarAnsatz}
\Phi &= f(\rho) \, e^{iN\theta}\,,&
(A_0,A_\rho,A_\theta) &= (0,0,a_\theta(\rho))\,,
\end{align}
where we chose the temporal gauge $A_0 = 0$ and the radial gauge $A_\rho = 0$. By the principle of symmetric critically, this configuration will also be a static solution of the full field equation. The axially symmetric ansatz reduces the field equations to,
\begin{align}
f'' + \frac{1}{\rho}f' - \frac{1}{\rho^2} f(1-a_\theta) - \frac{\lambda}{2}  f(f^2 - 1) &= 0\,,\label{eq:eom_radial1}\\
a_\theta'' - \frac{1}{\rho}a_\theta' + (1-a_\theta)f^2 &= 0\,.
\label{eq:eom_radial2}
\end{align}
Regularity gives us the profiles at the origin $f(0) = 0$ and $a(0) = 0$ while the boundary conditions above are now 
$$
\lim\limits_{\rho\rightarrow\infty} f(\rho)=1 \quad {\mathrm{and}} \quad 
\lim\limits_{\rho\rightarrow\infty} a_{\theta}(\rho)=N\,.
$$ 
The coupled system \eqref{eq:eom_radial1} -- \eqref{eq:eom_radial2} is non-linear and must be solved numerically, which is done using gradient flow with $4^{\mathrm{th}}$-order finite difference for derivatives to minimise the energy.

Although there is no known analytic solution, we can study the asymptotic form of the solutions for both $\rho \sim 0$ and $\rho \rightarrow \infty$. First, we will consider $f$ and $a_\theta$ near the origin, which admits the expansion $f(\rho) = \rho^N F(\rho^2)$ and $a_\theta(\rho) = \rho^2 \, G(\rho^2)$ where $F$ and $G$ are power series in $\rho^2$ with a non-zero coefficient for the leading term. Hence, we can write any general cylindrically symmetric solution of degree $N$ in the form,
\begin{equation}
\begin{split}
	&\Phi = \left(x_1+ix_2\right)^NF(x_1^2+x_2^2)\,,\\
	&A_{\mu} = (A_0, A_1, A_2) = \begin{pmatrix}0 \\ -x_2\, G(x_1^2 + x_2^2) \\ x_1\, G(x_1^2 + x_2^2) \end{pmatrix}\,,
\end{split}
 \label{Ansatz}
\end{equation}
where $F(\rho^2)$ and $G(\rho^2)$ are now nonlinear functions across the whole space, but can be expanded as a power series near zero. This reduces the field equations to,
\begin{align}
8\rho^2 F'' + 16F' - \rho^2F^3 + F(1 - 2\rho^2G^2+4G) &= 0\,,\label{eq:eom_power1}\\
4\rho^2 G'' + 8G' + F^2(1 - \rho^2G) &= 0\,.
\label{eq:eom_power2}
\end{align}

To consider the tails of the solutions, we linearise the system \eqref{eq:eom_radial1}-\eqref{eq:eom_radial2} around the vacuum $(f,a) = (1,N)$ which produces a decoupled system of two ODEs which yield the solution,
\begin{align}
f(\rho) &\sim 1 - \frac{q}{2\pi} K_0(\sqrt{\lambda}\rho)\,,&
a_\theta(\rho) &\sim N - \frac{m}{2\pi} \rho \, K_1(\rho)\,.
\label{eq:linear}
\end{align}
We can now understand the long-range static inter-vortex forces by assuming that a vortex at long-range acts as a point source \cite{speight1997static}, each with an associated scalar charge $q$ and magnetic dipole moment $m$. 
These point sources must satisfy the linear differential equations with solutions given in \eqref{eq:linear}. This leads to the linear interaction energy of two well separated point sources as,
\begin{equation}
E_{\text{int}}(R) = - \frac{q}{2\pi}K_0(\sqrt{\lambda}R) + \frac{m}{2\pi} K_0(R)\,,
\end{equation}
where $R$ is the separation. The key result is that the contribution from the Higgs field interaction is negative while the magnetic contribution is positive. Hence, for $\lambda < 1$ the Higgs field dominates at long-range causing vortices to attract, while for $\lambda > 1$ the magnetic field dominates at long-range causing vortices to repel \cite{speight1997static}. However, we will focus on the critically coupled case $(\lambda = 1)$, where the contributions from the Higgs field and magnetic field cancel each other with $q=m$, leading to no long-range interaction between static vortices.

\section{\label{sec:normalModes} Normal modes}
In this section we will study the normal modes for an $N=1$ vortex. This was first studied for several values of $\lambda$ by Goodband and Hindmarsh in \cite{Hindmarsch}, and we will take a similar approach here. Recently, these modes have been studied in more detail using different methods for $\lambda = 1$ \cite{Alonso_Izquierdo_2016,alonso2023spectral} and all $\lambda$ \cite{TomMartinShortrange}. To proceed we consider perturbations of the fields $(\Phi,A)$ around the background of a static vortex solution, and hence consider the quantities,
\begin{align}
\Psi(x) &= \Phi(x) - \Phi_s(x)\,,& \chi^\mu = A^\mu - a^{\mu}_s(x)\,,
\end{align}
where $(\Phi_s(x),a^{\mu}_s(x))$ is the static solution of \eqref{eq:eom1}-\eqref{eq:eom2} for given $\lambda$. Hence, the system is close to the static vortex precisely when the perturbations $\Psi$ and $\chi^\mu$ are small. This gives a correction to the action of the form,
\begin{equation}
S = S(\Phi_s,a^{\mu}) + \epsilon^2 S_2 + \mathcal{O}(\epsilon^3)\,,
\label{eq:action}
\end{equation} 
where $\epsilon$ is an arbitrary constant such that $\epsilon \ll 1$, and,
\begin{align}
S_2 &= \frac{1}{2}\int\xi^{\dagger}\mathcal{D}\xi \ d^2 x\,,& \xi^{\dagger}(\mathbf{x}) &= (\chi^- e^{i\omega t}, \chi^+ e^{-i\omega t},\overline{\Psi} e^{i\omega t},\Psi e^{-i\omega t})\,,
\end{align}
where $\xi$ is a vector of the perturbations, $\omega$ is the angular frequency of the linear mode, and $t$ denotes time.
Note that the linear action term vanishes because ($\Phi,A_\mu)$ is a solution of \eqref{eq:eom1}-\eqref{eq:eom2}, and as $\epsilon$ is small we can neglect all terms higher than quadratic, leaving only linear corrections to the equations of motion. We have also expanded the gauge field in terms of the total angular momentum state such that,
\begin{align}
\label{eq:AnsatzA}
    a^1_s &= -\frac{\sin(\theta)}{\rho}a_{\theta}(\rho) = -\frac{1}{2}(a_s^+ + a_s^-)\,,& a_s^2 &= \frac{\cos(\theta)}{\rho}a_{\theta}(\rho) = \frac{1}{2i}(a_s^+-a_s^-)\,, 
\end{align}
where
\begin{align}
\label{eq:AnsatzA+-}
    a_s^+ &= \frac{ia_{\theta}(\rho)}{\rho}e^{-i\theta}\,,& a_s^- = -\frac{ia_{\theta}(\rho)}{\rho}e^{i\theta}\,,
\end{align}
and $a_\theta(\rho)$ is a radial profile function found by solving \eqref{eq:eom_radial2}. $(A_0,A_\rho,A_\theta) = (A_0,0,a_\theta(\rho))$,
where we chose the radial gauge $A_\rho = 0$. The perturbation on the gauge field $\chi^{\mu}$ behaves the same.
Then, the total fields are,
\begin{align}
\label{pertAns1}
        \Phi(x) &= \Phi_s(x) + \epsilon \Psi(x)\,e^{-i \omega t}\,, & 
        \overline{\Phi}(x) &= \overline{\Phi}_s(x) + \epsilon \Psi(x)\,e^{i \omega t}\,,\\
        A^+(x) &= a_s^+(x) + \epsilon \chi^+(x)\,e^{-i \omega t}\,, & 
        A^-(x) &= a_s^-(x) + \epsilon \chi^-(x) \,e^{i \omega t}\,,
\end{align}

In order to set up the eigenvalue problem for the perturbations, we seek to remove the linear derivative terms by choosing the background gauge condition \cite{Cheng:1984vwu},
\begin{equation}
    \partial_{\mu}\chi^{\mu} - (\overline{\Psi}\Phi_s - \overline{\Phi_s}\Psi) = 0\,.
\end{equation}
This gauge choice removes the gauge degrees of freedom. Moreover, the Lorenz gauge $\partial_\mu A^{\mu} = 0$ is satisfied by this gauge condition. The Lorenz gauge is chosen for the full field theory dynamics because we found it to be the most suitable gauge choice for numerical simulations in Section \ref{sec:numerics}.

With the above ansatz, we obtain the eigenvalue equation from $\mathcal{D}$ by separating the time derivatives,
\begin{equation}
    \label{eq:eig}
    \mathcal{D}_{LG}\begin{pmatrix}
         \chi_+ \\ \chi_- \\ \Psi \\ \overline{\Psi}
    \end{pmatrix}  = \omega^2 \begin{pmatrix}
         \chi_+ \\ \chi_- \\ \Psi \\ \overline{\Psi}
    \end{pmatrix}\,,
\end{equation}
where 
$$
    \mathcal{D}_{LG} = \begin{pmatrix}
        D_1 & 0 & A & B\\
        0 & D_1 & C & E\\
        E & B & D_2 & V_1 \\
        C & A & V_2 & D_3 \\
    \end{pmatrix}\,,
$$
and  
\begin{align}
  \begin{split}\label{mylabel}
    &{}D_1 = -\Delta + |\Phi_s|^2\,, {}\\
    &{}D_2 = -\Delta - i(a_s^+ + a_s^-)\partial_x + (a_s^+ - a_s^+)\partial_y + \frac{\lambda}{2}(2|\Phi_s|^2-1)+a_s^+a_s^- + |\Phi_s|^2\,,{}\\
    &{}D_3 = -\Delta + i(a_s^+ + a_s^-)\partial_x - (a_s^+ - a_s^-)\partial_y + \frac{\lambda}{2}(2|\Phi_s|^2-1)+a_s^+a_s^- + |\Phi_s|^2\,,{}\\
         &
         \!\begin{alignedat}[t]{2}
           \hspace{0.67em}A &= i\overline{\partial_x\Phi_s} + \overline{\partial_y\Phi_s} + \overline{\Phi_s}a_s^+\,,\hspace{3em} 
    &&B = -i\partial_x\Phi_s-\partial_y\Phi_s + \Phi_s a_s^+\,,{}\\
    C &= i\overline{\partial_x\Phi_s}-\overline{\partial_y\Phi_s} + \overline{\Phi_s}a_s^-\,, 
    &&E = -i\partial_x\Phi_s + \partial_y\Phi_s + \Phi a_s^-\,,{}\\
    V_1 &= \frac{\lambda}{2}\Phi_s^2 - \Phi_s^2\,,
    &&V_2 = \frac{\lambda}{2}\overline{\Phi_s}^2 - \overline{\Phi_s}^2\,.{}\\
         \end{alignedat}\\
  \end{split}
\end{align}

The perturbations are given by \cite{Hindmarsch}
\begin{align}
\label{eq:pertAns}
    \Psi &= \sum\limits_k s_k(\rho) e^{i(N+k)\theta}\,,& \overline{\Psi}&= \sum\limits_k s_{-k}^*(\rho)e^{-i(N-k)\theta}\,,\\
    \chi^+ &= \sum\limits_k i\alpha_k(\rho) e^{i(k-1)\theta}\,, & \chi^- &= -\sum\limits_k i\alpha_{-k}^*(\rho) e^{i(k+1)\theta}\,,
\end{align}
where $N$ is the topological charge, and $k \in \mathbb{N}$ is the wave number.

Substituting the ansatz \eqref{eq:pertAns} for the perturbations, \eqref{eq:AnsatzA+-} for $a_s^{\pm}$ and \eqref{eq:polarAnsatz} for $\Phi_s$, we can reduce the eigenvalue problem \eqref{eq:eig} to a $1-$dimensional problem. For the case of this paper, we are only interested in the $N=1$, $k=0$ linear mode, in which case $s_0 = s_{-0}^*$ and $\alpha_0 = \alpha_{-0}^*$. Hence, for $N=1$ and $k = 0$,  the ansatz \eqref{eq:pertAns} simplifies to 
\begin{align}
    \psi_1(x) &= \cos{(\theta)}\ s_0(\rho)\,,\\
    \psi_2(x) &= \sin{(\theta)}\ s_0(\rho)\,,\\
    \chi_1(x) &= -\sin{(\theta)}\ \alpha_0(\rho)\,,\\
    \chi_2(x) &= \cos{(\theta)}\ \alpha_0(\rho)\,.
\end{align}
Hence, the eigenvalue problem \eqref{eq:eig} becomes,
\begin{equation}
    \begin{pmatrix}
        -\left(\partial_{\rho\rho}+\frac{1}{\rho}\partial_{\rho}\right) + f^2 + \frac{1}{\rho^2} & 
        \frac{2f}{\rho}\left(a_{\theta}-1\right)\\
        \frac{2f}{\rho}\left(a_{\theta}-1\right) &
        -\left(\partial_{\rho\rho}+\frac{1}{\rho}\partial_{\rho}\right) + 
        \frac{\lambda}{2}\left(3f^2-1\right)+\frac{1}{\rho^2}\left(a_{\theta}-1\right)^2\\
    \end{pmatrix}\begin{pmatrix}
        \alpha_0 \\ s_0
    \end{pmatrix} = \omega^2\begin{pmatrix}
        \alpha_0 \\ s_0
    \end{pmatrix}\,,
    \label{Eq:Eigenfunction}
\end{equation}
where we have eigenfunctions of the form $\xi = (\alpha_0,\alpha_0,s_0,s_0),$ and the system \eqref{eq:eig} has decoupled into two copies of equation \eqref{Eq:Eigenfunction}.

\begin{figure}
    \centering
    \includegraphics[width = 0.5\textwidth]{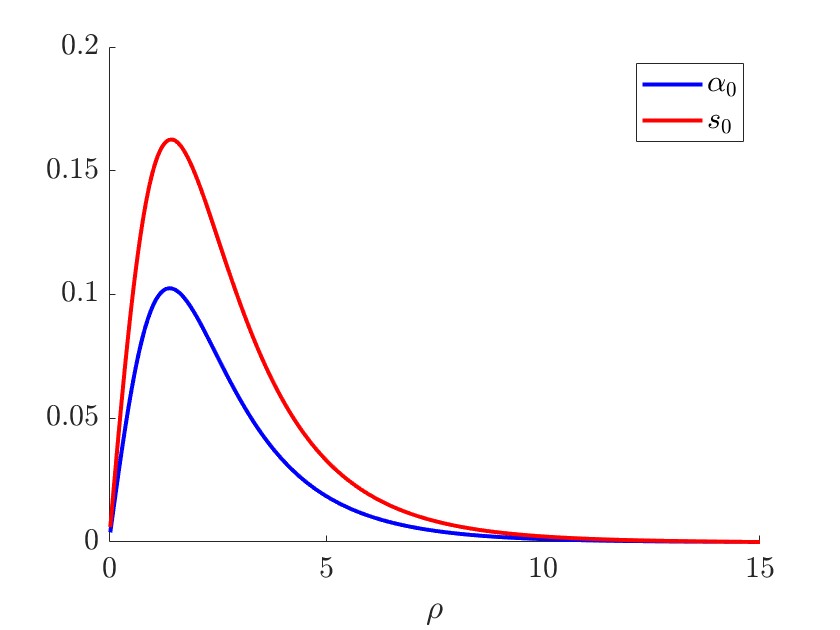}
    \caption{Plot of the solutions to \eqref{Eq:Eigenfunction}, normalised using the $\mathrm{L}_2$ norm \eqref{eq:L2_norm}, for the $N=1$ vortex shape mode. Here $\alpha_0(\rho)$ and $s_0(\rho)$ are the radial profile functions of the gauge field and the Higgs field, respectively.}
    \label{fig:Eigenfunction}
\end{figure}

We now employ a  central $2^{\rm nd}$-order finite-difference scheme to discretise the system of coupled ODE's \eqref{Eq:Eigenfunction}, and write the eigenvalue problem as a $2 \times 2$ block matrix, with entries of size $M \times M$. We then use MATLAB to find the eigenvalues of the block matrix. Boundary conditions are discussed in \cite{Alonso_Izquierdo_2016}.
We find that for $N=1$, we have only one normal mode, denoted the shape mode, as it is a radially symmetric mode that causes fluctuations in gauge-invariant quantities. We find that the mode has the frequency $\omega^2 = 0.77747$ and plot the eigenfunctions in \figref{fig:Eigenfunction}. We have normalised the eigenfunctions using the $\mathrm{L}_2$ norm,
\begin{equation}
    2\pi\int_0^{\infty} (\alpha_0(\rho)^2 + s_0(\rho)^2)\rho \ d\rho = 1\,.
    \label{eq:L2_norm}
\end{equation}

\section{\label{sec:numerics} Nonlinear numerical methods}

We seek dynamic solutions of the equations of motion \eqref{eq:eom1} and \eqref{eq:eom2}, which we find by numerically evolving the equations of motion from an initial condition of well separated Lorentz boosted vortices. We discretise the fields on a regular grid of $n_1 \times n_2$ lattice sites with spacing $h > 0$, where the discretised configuration space is the manifold $\mathcal{C} = (\mathbb{C}\times \mathbb{R}^3)^{n_1 n_2} \approx \mathbb{R}^{5 n_1 n_2}.$ We approximate the $1^{\rm st}$ and $2^{\rm nd}$ order spatial derivatives using central $4^{\rm th}$ order finite difference operators, yielding a discrete approximation $L_{dis}$ to the functional $L[\Phi,A]$ in \eqref{eq:Lagrangian}. We then evolved the discretised fields using a $2^{\rm nd}$ order Leapfrog method with time step $dt = 0.01$. Typical values used were $n_1 = n_2 = 601$ and $h = 0.1.$ 

Natural boundary conditions (detailed in Appendix \ref{boundaryconditions}) have been imposed to allow the phase to wind around the boundary as the fields evolve. Moreover, since we use a large amplitude in exciting the vortices, the dynamical solution exhibit radiation. 
Therefore, we have implemented damping boundary conditions near the boundary. We subtract the $1^{\rm st}$ order time-derivatives of the fields orthogonal to the gauge orbit, multiplied by a function, $K(x)$ from the equations of motion. $K(x)$ has boundary conditions $K(0) = 0$ and $K(\infty) = 1$, and is of the form
\begin{equation}
    K(x) = 1-(1-e^{\alpha (|x_1|-x_1^b)^2})(1-e^{\alpha (|x_2|-x_2^b)^2})\,,
\end{equation}
where $x_i^b$ is the location of the boundary. The constant $\alpha$ is chosen so that the damping boundary conditions only use $10\%$ of the boundary. Although the natural boundary conditions should allow for the radiation to pass through the boundary, some is reflected. The damping boundary conditions ensure that most of the radiation is absorbed so that it is not reflected back towards the bulk, affecting the behaviour of the mode in which we are interested. Note that the damping boundary conditions are not perfect and not all radiation is absorbed. To provide numerically accurate results, we altered the boundary conditions for a lattice of size $601\times601$ by varying the constant $\alpha$, so that the solution not only matches that of a solution found in a lattice of size $2001\times2001$, whereby the grid is sufficiently large that the radiation takes a long time to return to the system, but also fine-tuned the boundary conditions by choosing the best $\alpha$ so that there is as little radiation as possible.

During the development of the numerics, we considered other gauge choices motivated by work on vortex scattering \cite{ashcroft2017topological,moriarty1988dynamical,shellard1988vortex,Matzner}. Namely the temporal gauge $A_0 = 0$, which can be achieved either by using a gauge transformation to impose this condition after the boost or by imposing $A_1 = 0$ via a gauge transformation before a Lorentz boost. This gauge choice however is not compatible with the natural boundary conditions, as the temporal gauge provides no equation for the derivatives of the gauge field on the boundary. Hence, we chose the Lorenz gauge $\partial_{\mu}A^{\mu} = 0$, as it is compatible with the natural boundary conditions (detailed in Appendix \ref{boundaryconditions}). Note that we can check the numerics for a static vortex by checking that the gauge invariant quantities remain consistent no matter the gauge choice. 

The Leapfrog method was chosen to simulate numerical time integration for many reasons. Firstly, for symplecticity, meaning that geometric properties are conserved such as the total energy or magnetic flux. This provides advantages in maintaining stability and accuracy during large integration times. The method is also of $2^{\rm nd}$ order. As such, it is more efficient in terms of computational resources, which is essential in our case due to the volume of results. 

The $4^{\mathrm{th}}$-order Runge-Kutta method generally offers higher accuracy. However it is more computationally taxing, making it less suitable in our specific case. An artefact of the simplicity of the Leapfrog method is that it is more easily implemented compared to higher order methods such as RK$4$ and the Yoshida-Leapfrog method. To confirm that the Leapfrog method provides suitable accuracy, we also simulated solutions using the named higher order methods and found that the general structure of the phase space was the same. Even though individual simulations varied due to the chaotic nature of the solutions, as well as the differences in the integration techniques. Moreover, we tracked gauge-invariant quantities such as total energy and magnetic flux, and tracked the separation of the zeros of the Higgs field to ensure that the methods were consistent with each other.

We now explore a single vortex solution to the static equations of motion with excited shape mode. We can hence generalise an initial configuration for the vortex fields when the shape mode is excited.
\begin{equation}
    \begin{split}
        \phi_1(t,x_1,x_2) &= \mathcal{R}((x_1+ix_2)^N) F(x_1^2+x_2^2) + \epsilon \psi_1(x) \cos{(\omega t - \sigma(0))}\,,\\
        \phi_2(t,x_1,x_2) &= \mathcal{I}((x_1+ix_2)^N) F(x_1^2+x_2^2) + \epsilon \psi_2(x) \cos{(\omega t - \sigma(0))}\,,\\
        A_{\mu}(t,x_1,x_2) &= \begin{pmatrix}
            0 \\
            -x_2 G(x_1^2+x_2^2) + \epsilon \chi_1(x) \cos{(\omega t - \sigma(0))}\\
            x_1 G(x_1^2+x_2^2) + \epsilon \chi_2(x) \cos{(\omega t - \sigma(0))}\\
        \end{pmatrix}\,,
    \end{split}
\end{equation}
where $\sigma(0)$ is the initial phase of the mode, $\psi_i$, $\chi_i$ are the perturbations, $F$ and $G$ are the solutions of \eqref{eq:eom_power1} and \eqref{eq:eom_power2} respectively,  $\omega$ is the angular frequency and $\epsilon$ is the magnitude of the perturbation.

\begin{figure}
\centering
    \centering
    \includegraphics[width=0.8\textwidth]{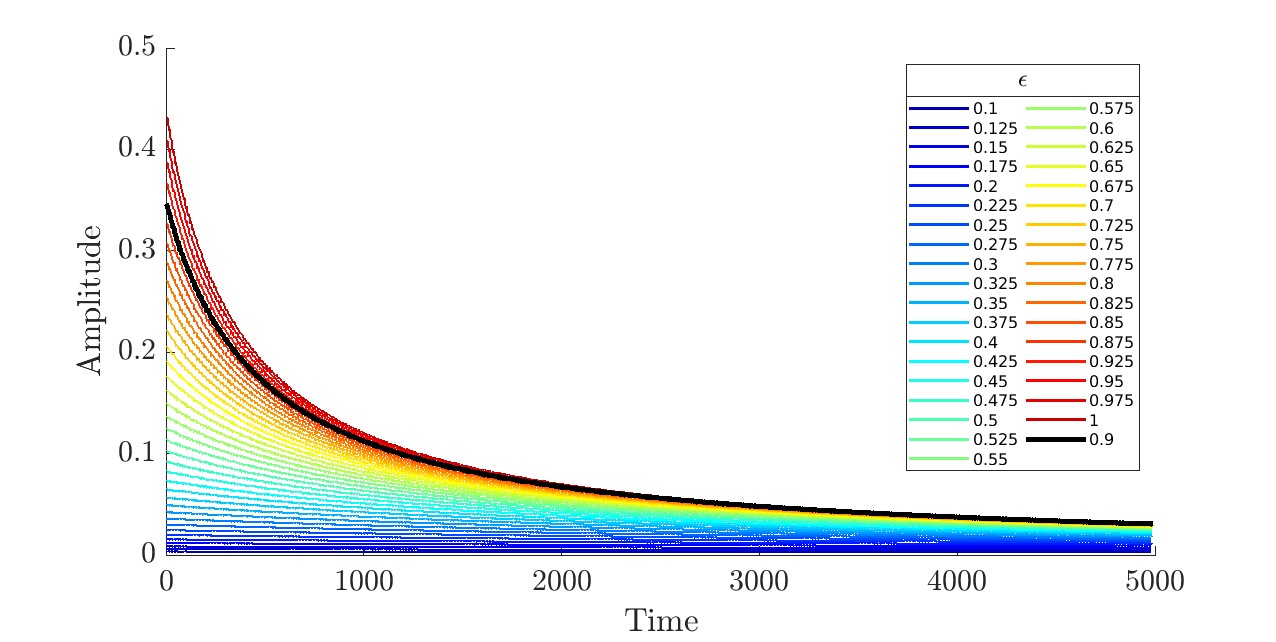}
    \caption{Change in amplitude of the $N=1$ vortex shape mode against time, where the amplitude is the magnitude of the fluctuations in the static potential energy. The black line with $\epsilon = 0.9$ corresponding to $A(0) = 0.317$ is our default initial amplitude in section \ref{sec:results}.}
    \label{fig:decayA}
\end{figure}

\begin{figure}
    \centering
    \includegraphics[width=0.8\textwidth]{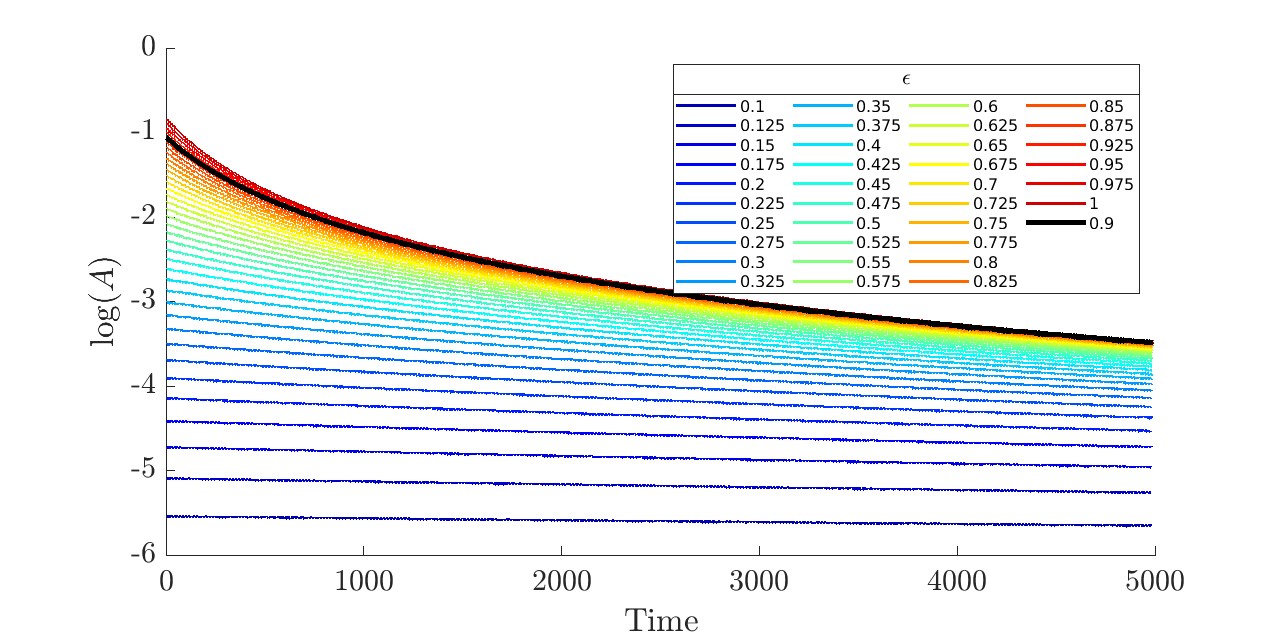}
    \caption{Log-plot of the amplitude of the $N=1$ shape mode against time, to show the exponential decay of the excitation. The black line with $\epsilon = 0.9$ corresponding to $A(0) = 0.317$ is our default initial amplitude in section \ref{sec:results}. }
    \label{fig:decayB}
\end{figure}

We can now simulate a single vortex of degree $N$, with excited shape mode. We can hence study the amplitude of the $N = 1$ excitation over time, by calculating the amplitude of the static potential energy, see \figref{fig:decayA}. \figref{fig:decayA} shows how the amplitude of the excitation changes with time. Note that the energy is conserved. However the damping boundary conditions remove radiation from the system, hence the energy is allowed to decrease. The solid black line indicates the choice of $\epsilon$ used for the majority of our results. We denote the initial amplitude of the excitation $A(0)$, where \begin{equation}A(0) = \frac{1}{2}(\epsilon\omega)^2\,.\label{eq:amplitudeEps}\end{equation} We can see that there is an exponential decay by taking a logarithm of the amplitude, see \figref{fig:decayB}, whereby for $\epsilon < 0.7$, the resulting curves are straight lines. Initially, we see that for larger $\epsilon$, the amplitude of the shape mode decays faster. However, changing the initial amplitude is the same as shifting through time, see \figref{fig:decayA}, whereby we can shift along the Time-axis such that all different initial amplitudes can be considered as a decayed amplitude along the same curve.


\figref{fig:decayA} gives us a range of suitable amplitudes for the excitation of the shape mode to scatter excited vortices. To simulate scattering, we must first impose an initial velocity for the vortex configuration so that the vortex moves. We then boost the vortex using a Lorentz transformation.

Our coordinates transform as $\hat{x} = \Lambda\tilde{x}\,$ where
\begin{align}
\Lambda &= \begin{pmatrix}\gamma & -\gamma v & 0 \\ -\gamma v & \gamma & 0 \\ 0 & 0 & 1\end{pmatrix}\,, &  \Lambda^{-1} &= \begin{pmatrix}\gamma & \gamma v & 0 \\ \gamma v & \gamma & 0 \\ 0 & 0 & 1\end{pmatrix}\,,
\end{align} 
and $\gamma = \frac{1}{\sqrt{1-v^2}}$ is the Lorentz factor. Consequentially, the Higgs field and gauge field then transform as, 
\begin{align}\Phi(x) &\mapsto \hat{\Phi}(x) = \Phi(\Lambda^{-1}x)\,,&
A_{\mu}(x) &\mapsto \hat{A_{\mu}}(x) = \Lambda A_{\mu}(\Lambda^{-1}x)\,.
\end{align}
Hence, we have an initial configuration for our two-dimensional dynamical numerics, detailing an axially symmetric vortex with an initial velocity and an excited shape mode.

\begin{equation}
\label{Ansatz2}
    \begin{split}
        \tilde{\phi_1}(t,x_1,x_2) &= \mathcal{R}((\gamma (x_1+v t)+ix_2)^N) F(\gamma^2(x_1+v t)^2+x_2^2) + \epsilon \psi_1(\tilde{x}) \cos{(\omega \gamma (t + v x_1) - \sigma(0))}\,,\\
        \tilde{\phi_2}(t,x_1,x_2) &= \mathcal{I}((\gamma (x_1+v t)+ix_2)^N) F(\gamma^2(x_1+v t)^2+x_2^2) + \epsilon \psi_2(\tilde{x}) \cos{(\omega \gamma (t + v x_1) - \sigma(0))}\,,\\
        \tilde{A_{\mu}}(t,x_1,x_2) &= \begin{pmatrix}
            -\gamma v x_2 (\gamma^2(x_1+v t)^2+x_2^2) + \gamma v \epsilon \chi_1(\tilde{x}) \cos{(\omega \gamma (t + v x_1) - \sigma(0))}\\
            -\gamma x_2 (\gamma^2(x_1+v t)^2+x_2^2) + \gamma \epsilon \chi_1(\tilde{x}) \cos{(\omega \gamma (t + v x_1) - \sigma(0))}\\
            \gamma (x_1 + v t) G(\gamma^2(x_1+v t)^2+x_2^2) + \epsilon \chi_2(\tilde{x}) \cos{(\omega \gamma (t + v x_1) - \sigma(0))}\\
        \end{pmatrix}\,.
    \end{split}
\end{equation}
For large initial amplitudes, the non-linear terms in \eqref{eq:action} become significant, and we observe that the energy is phase dependant, varying up to order $\mathcal{O}(\epsilon^3)$ for a $\pi-$shift. 

It is outlined in Appendix A how to excite the same mode using a Derrick's scaling. We find that the mode excitation can be well approximated by a scaling of the fields. However, this allows less freedom in the choice of the initial phase. To alter the phase using the Derrick's method, we must evolve the vortex in time to numerically change the initial phase of the mode, which results in a small decay in the energy. 
Using the method by which we alter the phase in the Derrick's approximation, we can also alter the phase the same way for the linearisation. By changing the phase this way, the amplitude of the shape mode decays by approximately $1e-4$, which is significantly less than the contribution to the energy of the higher order terms in the linearisation. Because of this, we will show in \cref{sec:results} the phase space plot for both methods.

The initial field configurations \eqref{Ansatz2} are solutions to the dynamic equations of motion \eqref{eq:eom1} and \eqref{eq:eom2} and can be used to simulate a single degree $N$ vortex with excited shape mode. We seek to study the scattering of these excited degree $N=1$ vortices; hence we must create multi-vortex field configurations that are also solutions to the equations of motion \eqref{eq:eom1} and \eqref{eq:eom2}. The Abikrosov ansatz \cite{abrikosov1957magnetic} allows us to find field configurations detailing well-separated Lorentz boosted vortices with excited shape modes. The Abikrosov ansatz for a given vortex solution $(\tilde{\Phi}(t,x),\tilde{A_{\mu}}(t,x))$ is
\begin{align}
\label{AbikEq}
\hat{\Phi} &= \prod_i\tilde{\Phi}(x - d_i)\,,&
\hat{A_{\mu}} &= \sum_i \tilde{A_{\mu}}(x - d_i)\,,
\end{align}
where $d_i$ are the positions of the vortex centres. The approximation works well when the vortices are well separated from each other, such that the separation is much larger that the vortex core size, namely, $2d_i \gg 1$.

\begin{figure}
\centering
    \includegraphics[width = 0.85\textwidth]{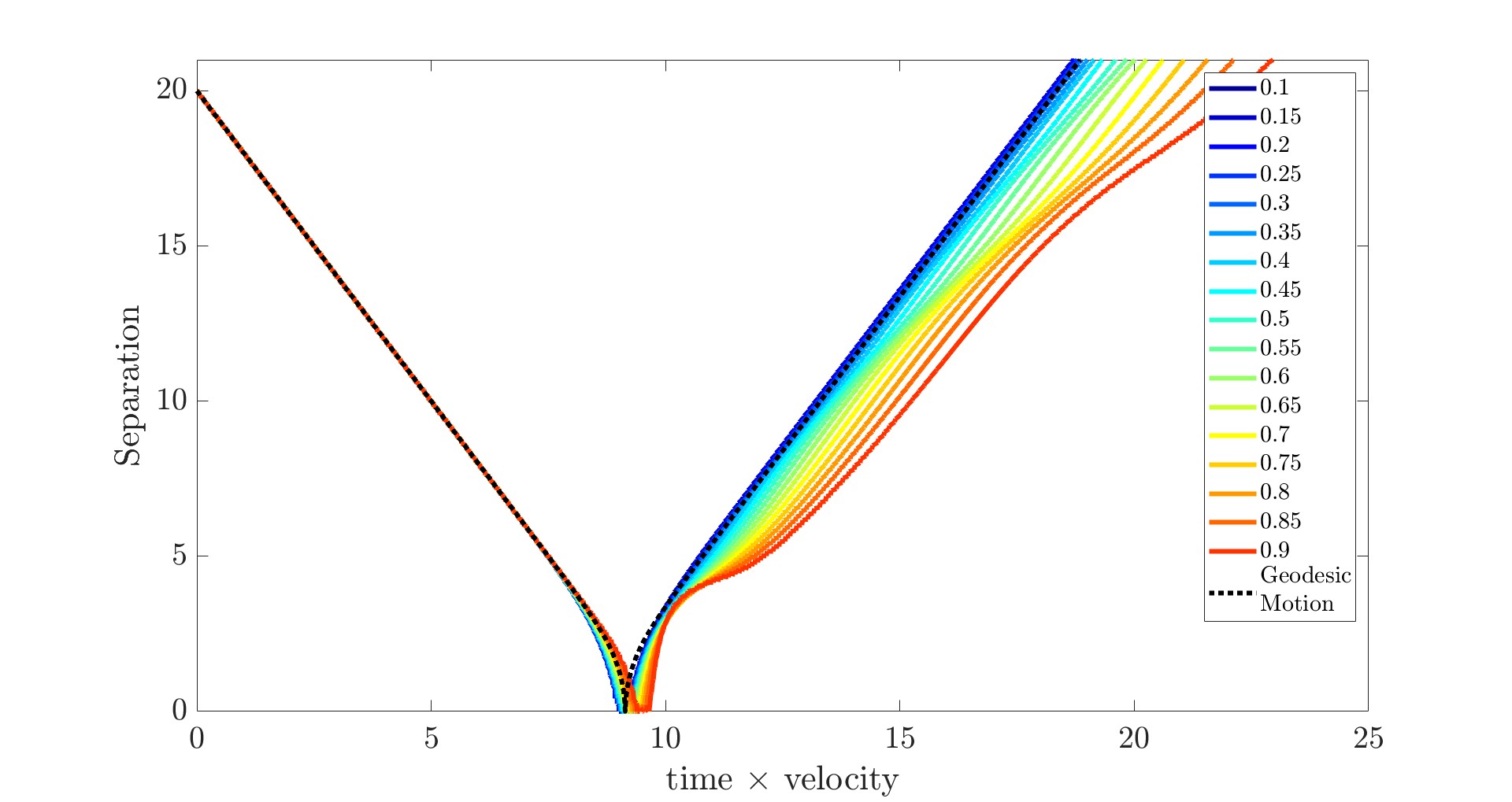}
    \caption{Separation of the zeros of the Higgs field (interpreted as the vortex position) of 2 vortices scattering at various initial velocities against time. Here the time is rescaled by the velocity such that $t \mapsto v_{\mathrm{in}} t$, where $v_{\mathrm{in}}$ is the initial velocity of the vortices.}
    \label{fig:crit_scat_rs}
\end{figure}

To confirm that our numerics are working correctly, we can simulate the scattering of vortices at critical coupling, using the configuration \eqref{Ansatz2}, and setting the perturbations to zero. We can then track the zeros of the condensate to plot the separation for a set of initial velocities. In the moduli space approximation, the trajectories are independent of the initial velocity. This leads us to rescale our trajectories as $t \to v_{\mathrm{in}} t$, where $v_{\mathrm{in}}$ is the initial velocity of the vortices. \figref{fig:crit_scat_rs} shows the scattering of two vortices of multiplicity one, with varied initial velocities. As expected, the trajectories initially lie on the same curve until $t \approx 8.$ For small velocities $v_{\mathrm{in}} < 0.3$, our numerics match the expected behaviour from the moduli space approximation (dotted line) whereby they travel with constant velocity and scatter at $90^{\circ}$. For larger velocities, the numerics deviates significantly from the moduli space approximation which is only valid for small velocities. For velocities close to one the trajectories show new kinds of behaviour which goes beyond the scope of this paper.

\section{\label{sec:results} Scattering of excited vortices}

In this section, we study the scattering behaviour of two $N=1$ critically coupled vortices with excited shape modes. The excitation leads to an interesting scattering behaviour dependent on initial velocity, as well as amplitude and phase of the shape mode. We look at snapshots of a numerical simulation which show the scattering of the excited vortices. We also plot different vortex trajectories, where we vary the initial phase of the shape mode. Furthermore, we show a plot summarising a sampling of scattering outcomes for a fixed amplitude, where we vary the initial velocity and the phase of the shape mode. We then discuss how this summary is different if we change the initial amplitude of the shape mode. Finally, we give a brief discussion regarding changing the relative phase of the shape mode between the two vortices.

For all simulations discussed in this section, the vortices are located at $d_i = \pm10$, where $d_i$ is defined in \eqref{AbikEq}. This separation was chosen so that the vortices are initially well separated so that the forces between them can be neglected. 
Unless stated otherwise, we consider solutions for a fixed initial amplitude $A(0) = 0.317$. This corresponds to $\epsilon = 0.9$, where $\epsilon$ is the magnitude of the perturbation, defined in \eqref{pertAns1}. The initial amplitude is calculated from $\epsilon$, see \eqref{eq:amplitudeEps}. We choose a sufficiently large initial amplitude $A(0)$ such that there is enough energy in the shape mode for a considerable amount of interesting behaviour in the excited-scattering process.
We label the initial phase of the shape mode by $\sigma(0) \in [0,2\pi)$, defined in \eqref{Ansatz2}. Unless stated otherwise, the two vortices are in phase with each other. We denote the initial velocity of the vortices by $v_{\rm in}$.
\begin{figure}
\centering
\begin{subfigure}[b]{0.23\textwidth}
\centering
\includegraphics[width=\textwidth]{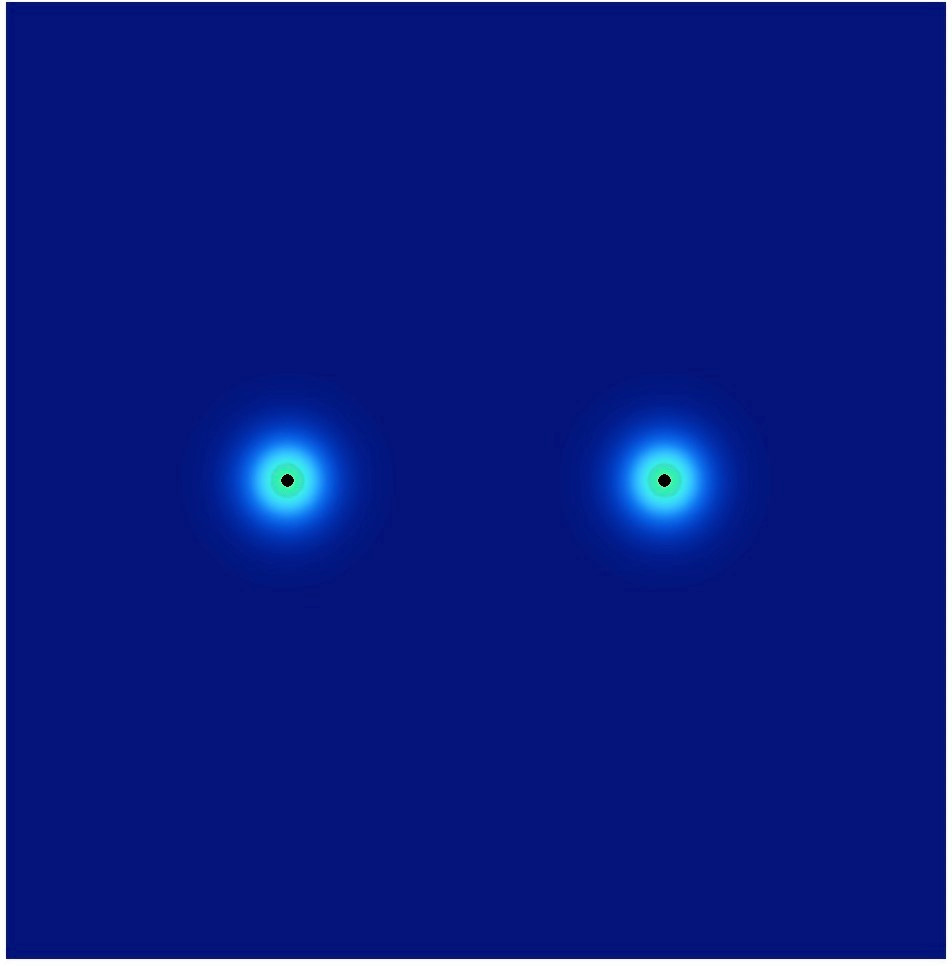}
\caption{$t = 0$}
\end{subfigure}
\begin{subfigure}[b]{0.23\textwidth}
\centering
\includegraphics[width=\textwidth]{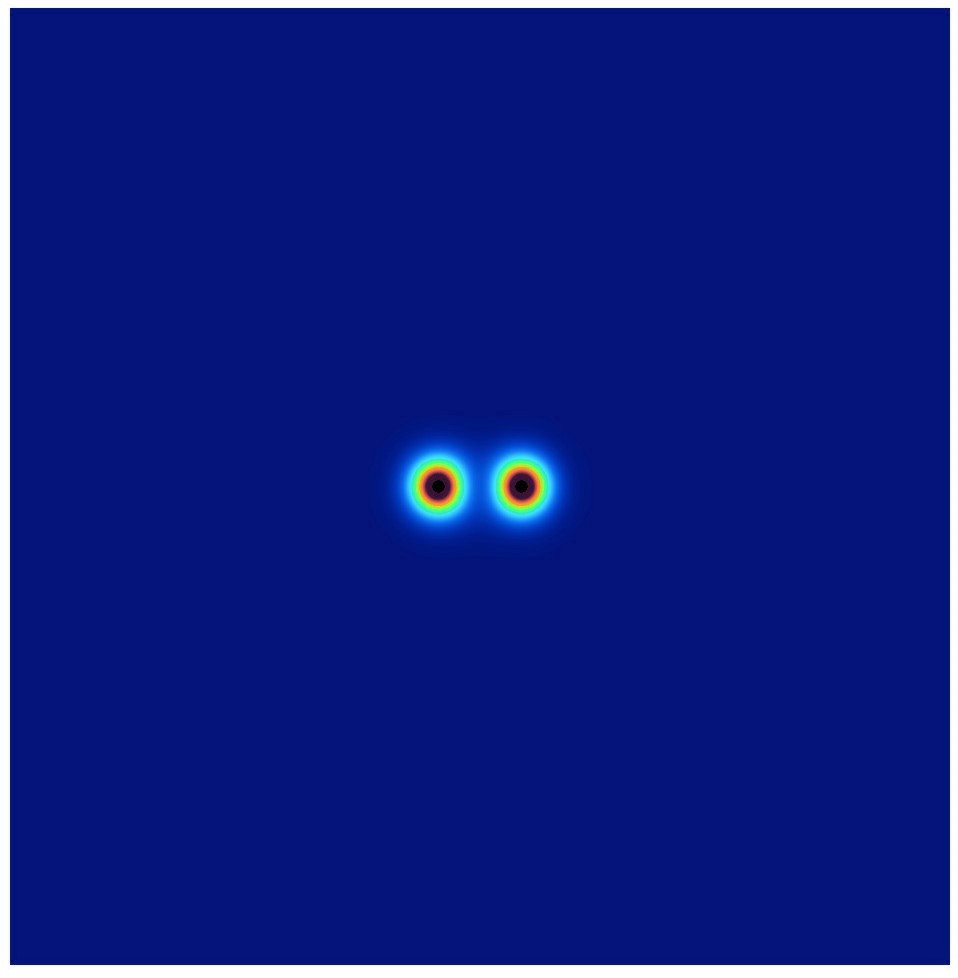}
\caption{$t = 270$}
\end{subfigure}
\begin{subfigure}[b]{0.23\textwidth}
\centering
\includegraphics[width=\textwidth]{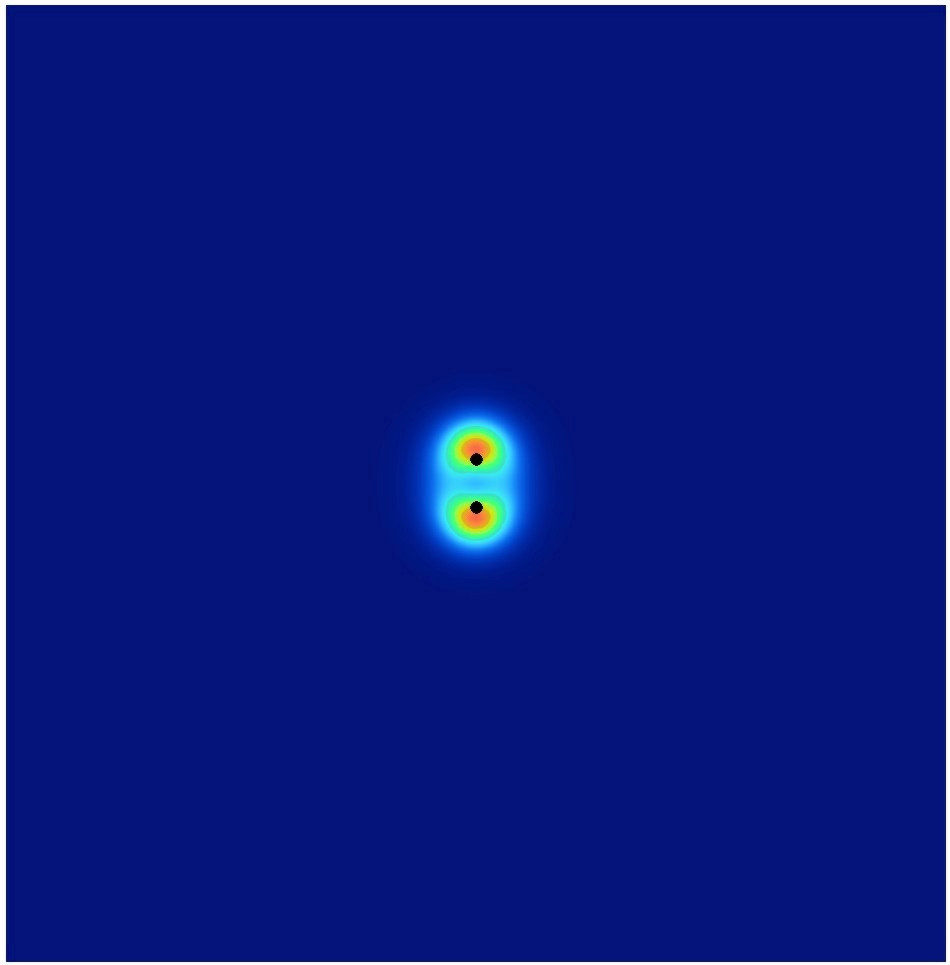}
\caption{$t = 280$}
\end{subfigure}
\begin{subfigure}[b]{0.23\textwidth}
\centering
\includegraphics[width=\textwidth]{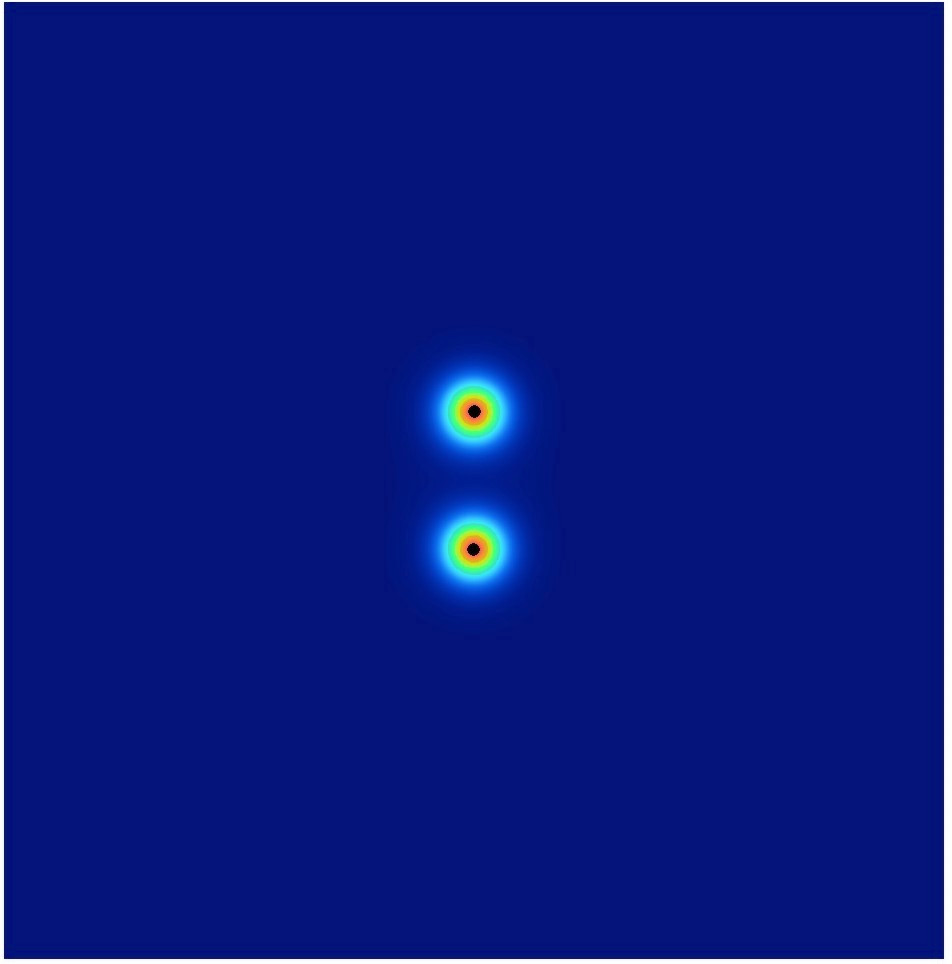}
\caption{$t = 300$}
\end{subfigure}
\begin{subfigure}[b]{0.05\textwidth}
\centering
\includegraphics[width=\textwidth]{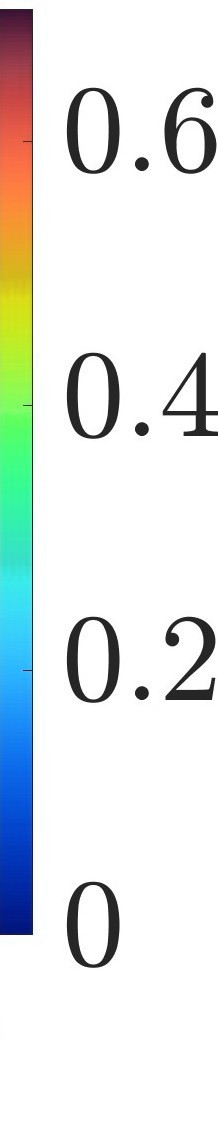}
\end{subfigure}
\begin{subfigure}[b]{0.23\textwidth}
\centering
\includegraphics[width=\textwidth]{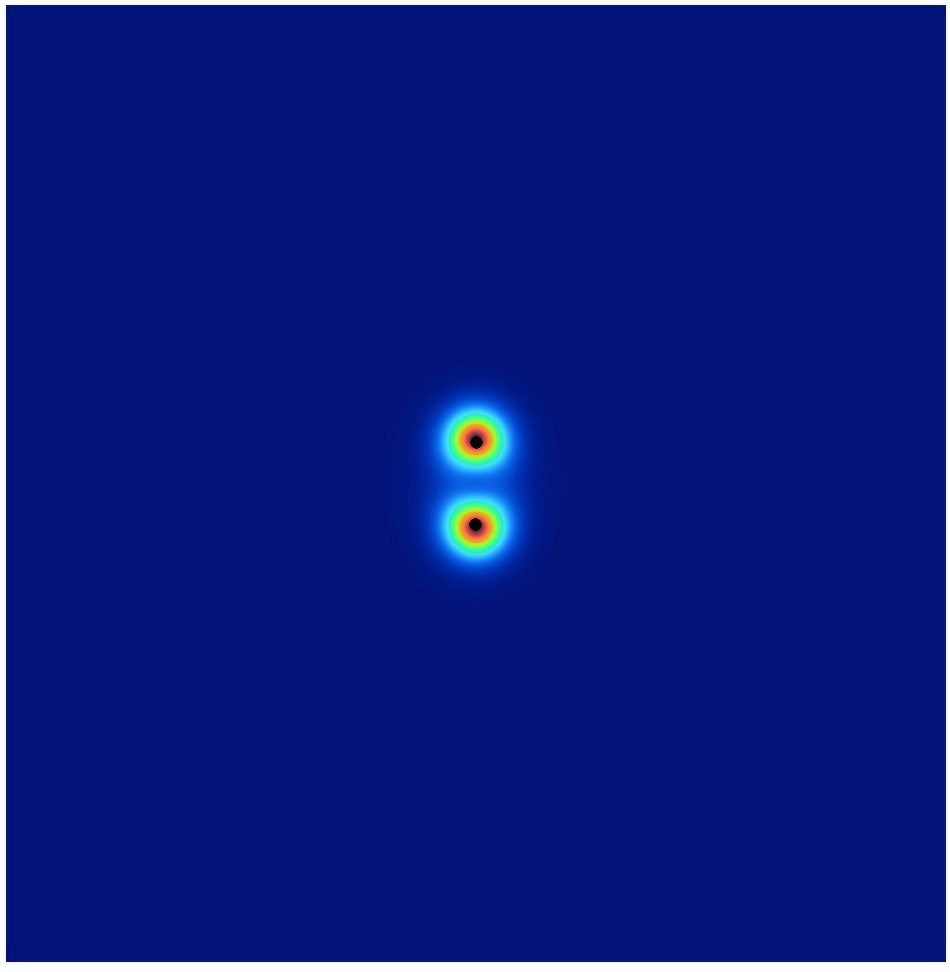}
\caption{$t = 360$}
\end{subfigure}
\begin{subfigure}[b]{0.23\textwidth}
\centering
\includegraphics[width=\textwidth]{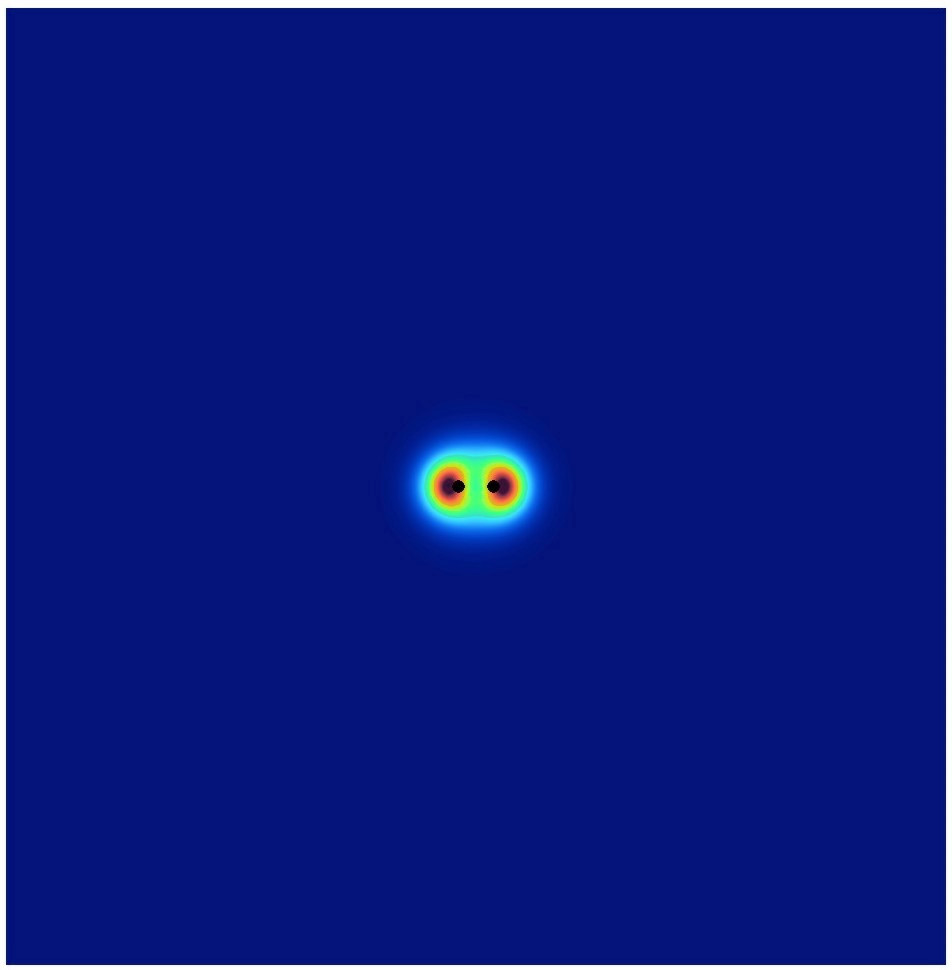}
\caption{$t = 370$}
\end{subfigure}
\begin{subfigure}[b]{0.23\textwidth}
\centering
\includegraphics[width=\textwidth]{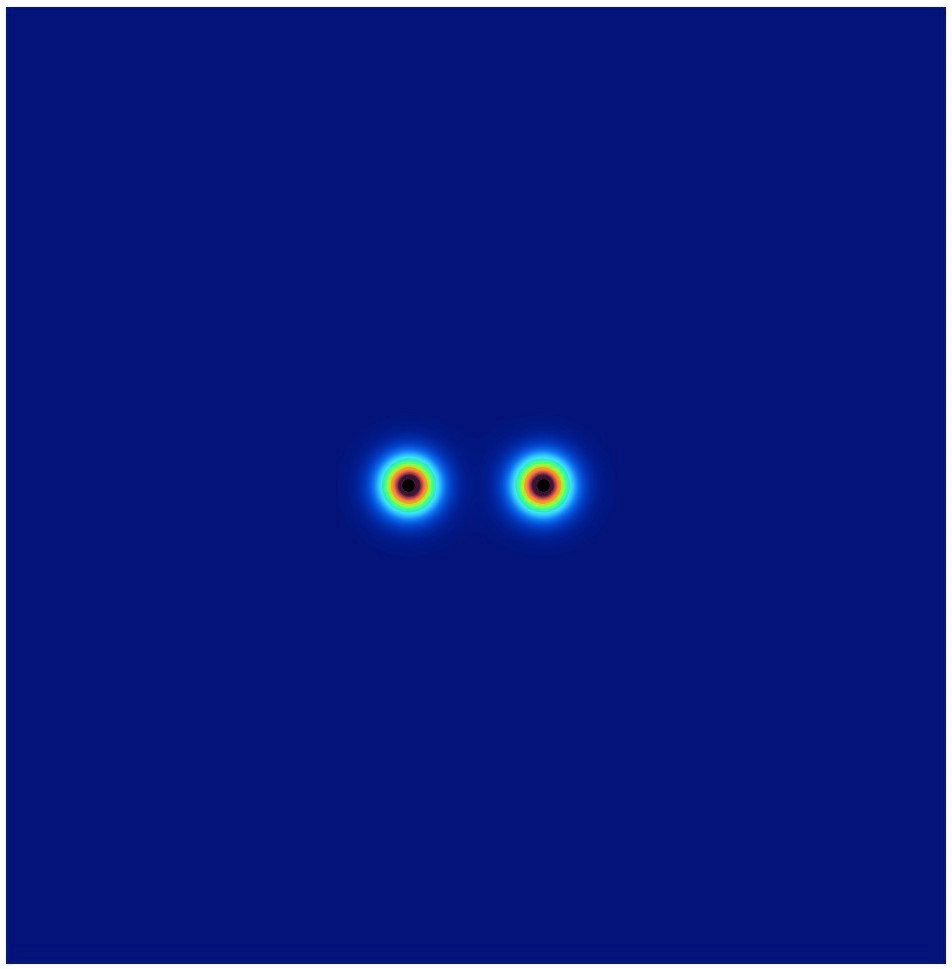}
\caption{$t = 400$}
\end{subfigure}
\begin{subfigure}[b]{0.23\textwidth}
\centering
\includegraphics[width=\textwidth]{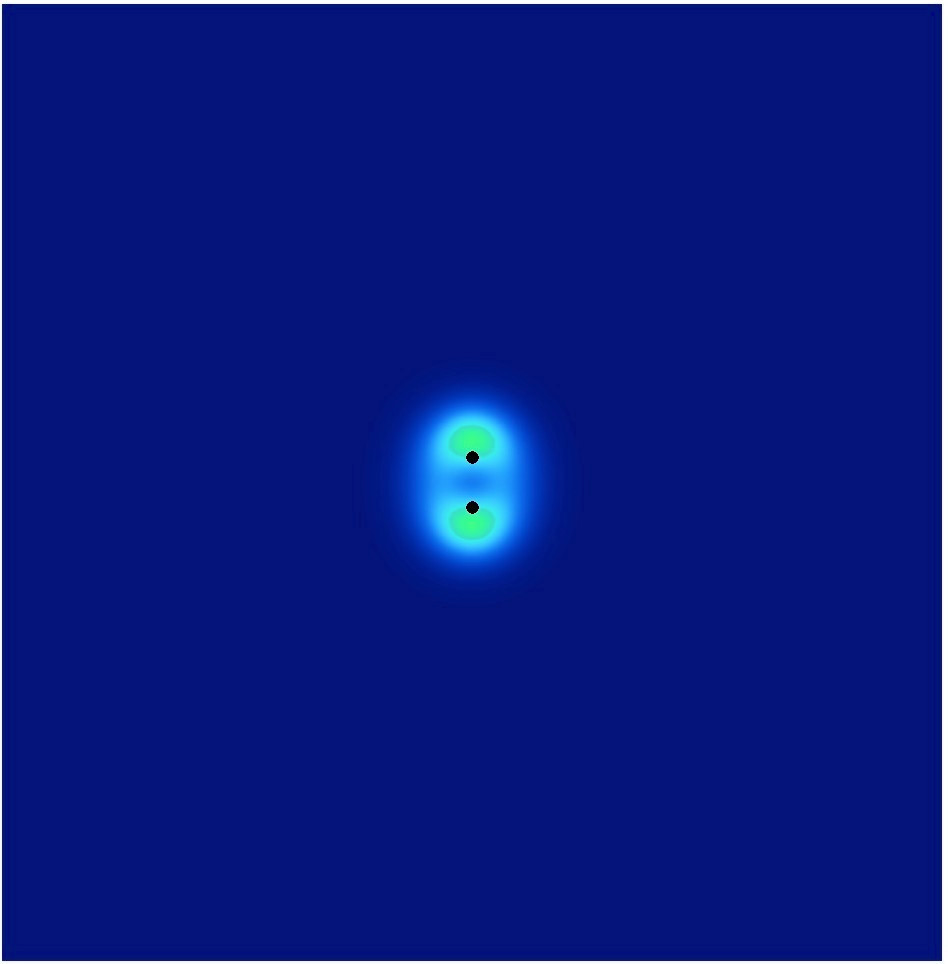}
\caption{$t = 450$}
\end{subfigure}
\begin{subfigure}[b]{0.05\textwidth}
\centering
\includegraphics[width=\textwidth]{Image_c.jpg}
\end{subfigure}
\caption{Heat plots of the energy density, showing snapshots through time of an excited vortex scattering, with initial phase $\sigma(0) = 2.2612$, initial velocity $v_{\mathrm{in}} = 0.01$, and initial separation $d = 20$. The black dots indicate the zeros of the Higgs field. This figure shows how the vortices accelerate towards each other and then scatter at $90^{\circ}$. The vortices then slow before accelerating towards each other and scattering at $90^{\circ}$ again, which repeats many times.}
\label{fig:snapshot}
\end{figure}

Firstly, we show snapshots of a simulation for a 2-vortex scattering with excited shape modes in \figref{fig:snapshot}. The initial phase of the shape mode for each vortex is $\sigma(0) = 2.2612$, and $v_{\rm in} = 0.01$. We display the energy density as a heat plot and overlay the zeros of the Higgs field as black dots. We can see that the energy density fluctuates as a result of the excited shape mode. At critical coupling, there are no static forces between vortices, and vortices scatter at right angles in agreement with the moduli space approximation. We find that this is no longer the case for excited vortices. We refer to this multi-bounce behaviour as a quasi-bound state.

\begin{figure}
    \centering
    \begin{subfigure}[b]{0.49\textwidth}
    \centering
        \includegraphics[width=\textwidth]{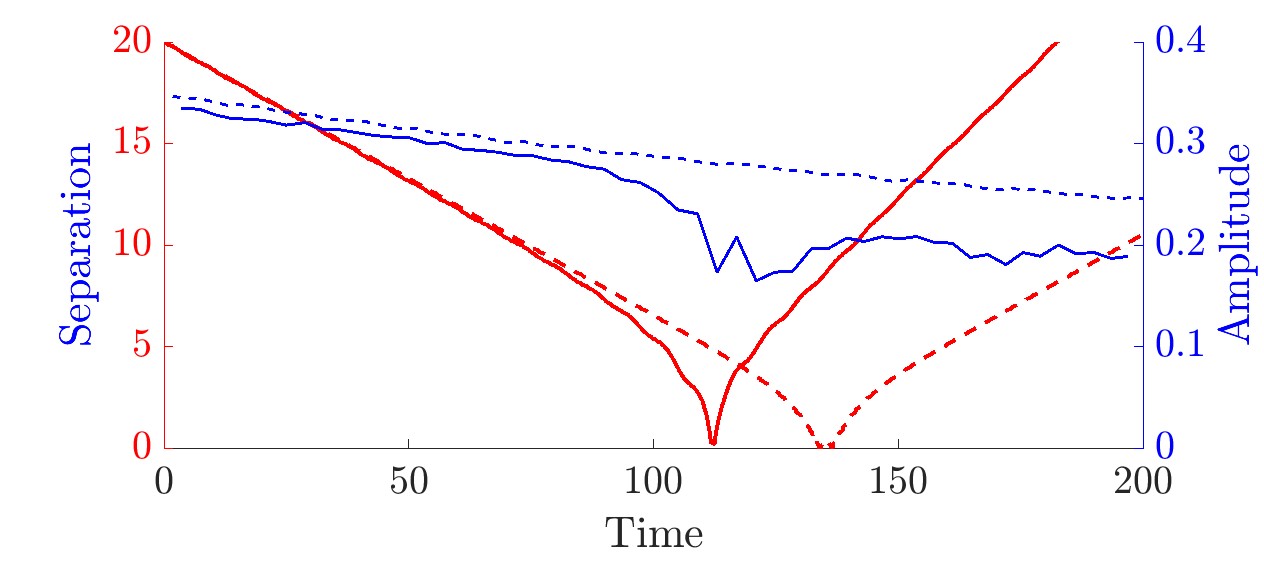}
        \caption{$\sigma(0) = 0$}
        \label{fig:track1}
    \end{subfigure}
    \centering
    \begin{subfigure}[b]{0.49\textwidth}
    \centering
        \includegraphics[width=\textwidth]{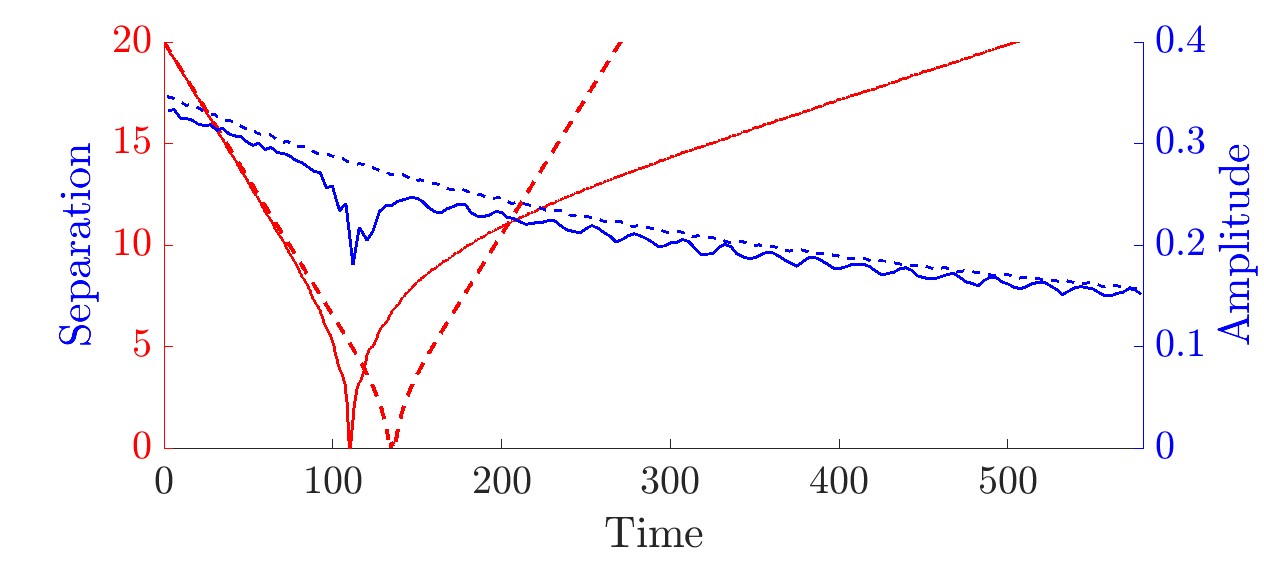}
        \caption{$\sigma(0) = \frac{13\pi}{16}$}
        \label{fig:track2}
    \end{subfigure}
    \centering
    \begin{subfigure}[b]{0.49\textwidth}
    \centering
        \includegraphics[width=\textwidth]{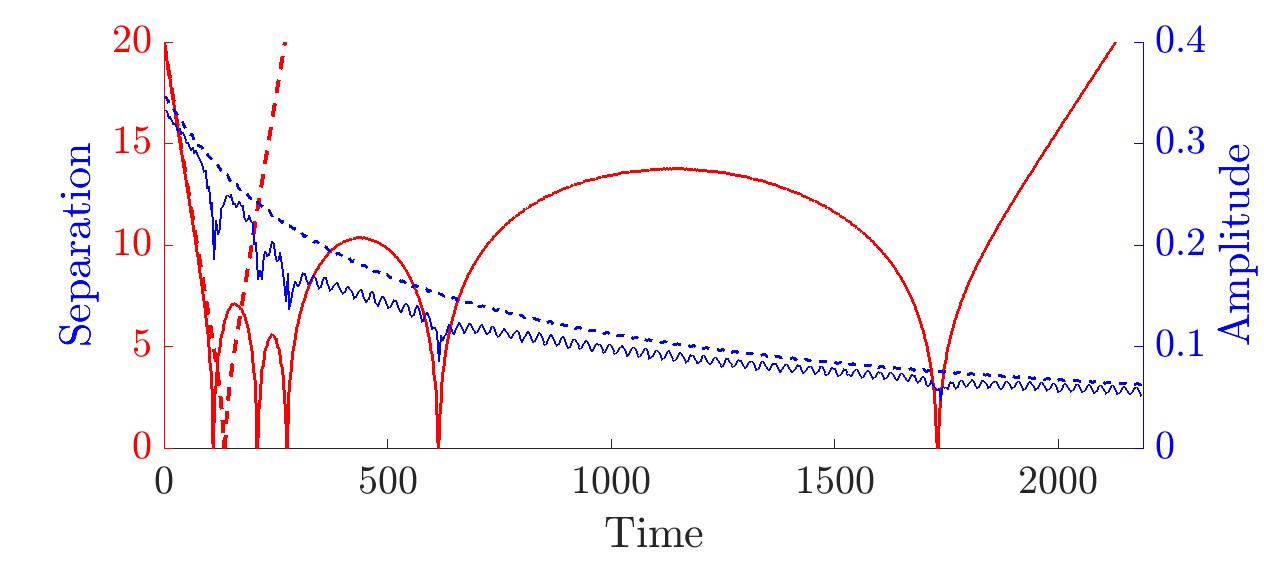}
        \caption{$\sigma(0) = \frac{15\pi}{16}$}
            \label{fig:track3}
    \end{subfigure}
    \centering
    \begin{subfigure}[b]{0.49\textwidth}
    \centering
        \includegraphics[width=\textwidth]{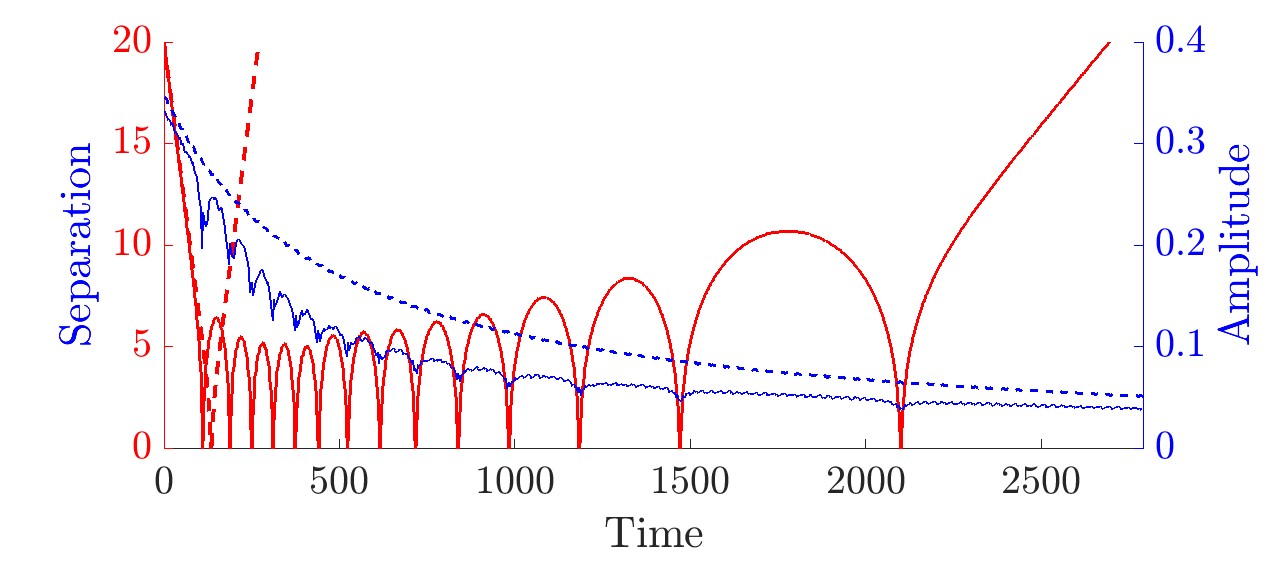}
        \caption{$\sigma(0) = \frac{17\pi}{16}$}
            \label{fig:track4}
    \end{subfigure}
    \centering
    \begin{subfigure}[b]{0.49\textwidth}
    \centering
        \includegraphics[width=\textwidth]{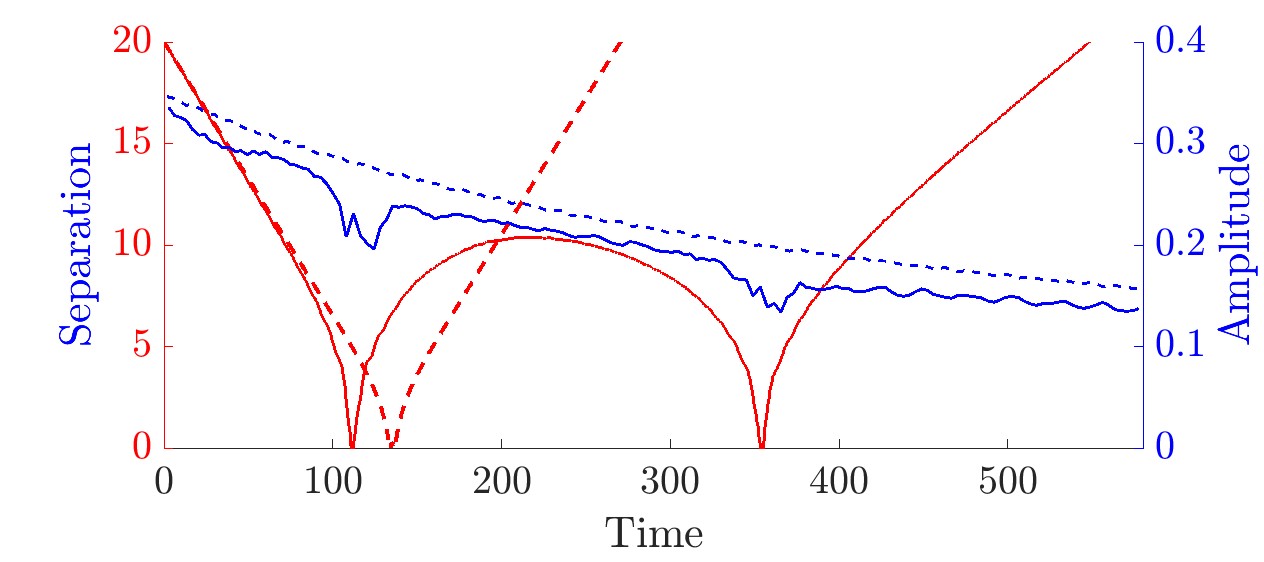}
        \caption{$\sigma(0) = \frac{5\pi}{4}$}
        \label{fig:track5}
    \end{subfigure}
    \centering
    \begin{subfigure}[b]{0.49\textwidth}
    \centering
        \includegraphics[width=\textwidth]{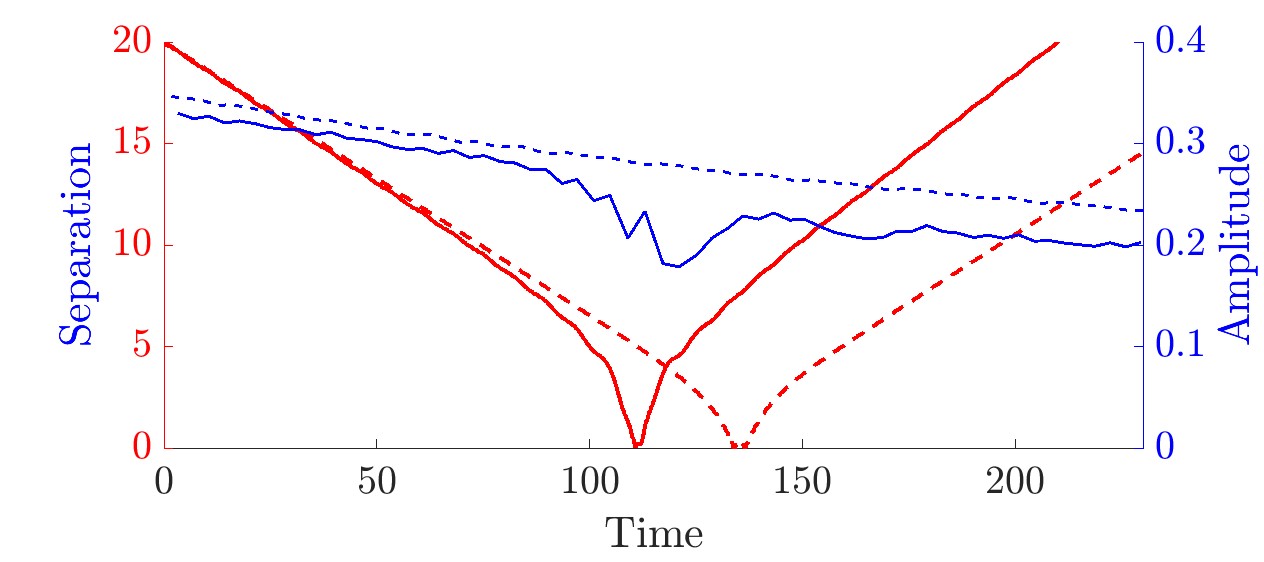}
        \caption{$\sigma(0) = \frac{25\pi}{16}$}
        \label{fig:track6}
    \end{subfigure}
    \caption{Tracking of separation of the vortices with time, plotted in red. Blue indicates the amplitude of the excitation per vortex. We show 6 plots, with different initial phases, and fixed initial velocity $v_{\mathrm{in}} = 0.06725$. The dashed red line indicates the standard scattering process with no excitation but with the same initial velocity. The dashed blue line indicates the amplitude of the excitation in the absence of the scattering.
    }
    \label{fig:tracking}
\end{figure}

For a fuller picture of the 2-vortex scattering, we can track the zeros of the Higgs field, as seen in \figref{fig:tracking}. We plot the separation of the zeros for a set of solutions to show the trajectories of the vortices as a function of time. We have only varied the initial phase $\sigma(0)$ for fixed velocity $v_{\rm in} = 0.06725$. The solid red line shows the separation, $d$, of the zeros of the Higgs field of the two vortices with excited shape mode, and the solid blue line shows is the amplitude of the excitation. The dashed red line shows the separation of two vortices with the same initial configuration, but no excitation. The dashed blue line indicates the amplitude of a single vortex with the same mode excitation. 

First, let us discuss the excited vortex scattering in general. We can see that the trajectories of the vortices with excited shape modes is different to that with no excitation. Initially, there is no deviation between the trajectories of the vortices with or without excitation. There is also no curvature in the trajectories before $d \approx 17$, showing that the vortices travel at a constant velocity initially. This is because the length scale of the mode is approximately the same as the size of the vortices, which fall off exponentially at approximately $l = 8.5 (d = 2l)$, see \figref{fig:Eigenfunction}. 

For $d < 17$, the trajectory of the excited vortices begins to deviate from that of the standard scattering. We observe an increasing slope in the trajectory of the excited vortices, and the excited vortices also collide sooner than with no excitation. We can hence see that the vortices begin to accelerate towards each other within this region. This interaction is similar to the behaviour of vortices in type \rom{1} superconductors, where vortices are attractive. 

We can see in \cref{fig:track1,fig:track2,fig:track6} scattering solutions where the vortices only scatter once. More than this, we can see in \cref{fig:track1,fig:track6} that the exit velocity $v_{\rm out}$ is greater than the initial velocity $v_{\rm in}$. We hence observe behaviours of type \rom{2} superconductors, as the vortices move apart from each other at a velocity greater than that initially configured.

We can conjecture that this attraction and repulsion is the result of the changing length scales of the vortex as it wobbles. We discussed in Appendix \ref{sec:appendixDerrick} that the interaction energy fluctuates between positive and negative but is mainly negative. We also see that the scalar interaction fluctuates with a magnitude higher than that of the magnetic interaction. Since the interaction energy is mainly negative, see \figref{fig:InteractionEnergy}, we can conjecture that the mode excitation results in an attraction where the scalar interaction dominates; however, since the interaction energy also fluctuates to positive, the mode excitation also produces a repulsive interaction, where the magnetic interaction dominates. This shows that geodesic motion is not the correct approximation for excited vortex scattering, as it does not explain attraction due to excitation.

In all tracking plots, the amplitude of the excitation drops after the vortices collide. We can see that this is a result of the collision as this is a deviation from the dashed blue line. This is because energy from the mode is transferred to the kinetic energy of the vortex. After the excited vortices scatter, the amplitude increases slightly, suggesting that kinetic energy from the vortices is transferred back to the excitation. If there is more energy in the excitation, the vortices become more attractive, and hence we observe that they scatter again. Near the end of the simulation, we can see that the amplitude of the excitation has decreased significantly, especially for \cref{fig:track3,fig:track4}. It is possible that there is not enough energy left in the excitation, as it radiates energy due to the fast decay of the amplitude. This means that not enough energy can be transferred to the kinetic energy, and hence the vortices escape.

There are some slight fluctuations in the amplitude after the vortices collide. We believe this fluctuation to be a result of the Doppler effect as radiation is emitted from the vortices as they travel, which we have reproduced by studying the Doppler effect.

\cref{fig:track3,fig:track4,fig:track5} display a quasi-bound state, where we have multiple bounces. \figref{fig:track5} shows a 2-bounce scattering solution, \figref{fig:track3} shows a 4-bounce solution, and \figref{fig:track4} shows a 13-bounce solution. We can see from the trajectories that the size of the bounce windows increases with time. This could be argued to be a result of the decay of the mode. As the mode decays, it loses energy, resulting in a reduced attractive quality as time progresses. This behaviour is expected as it is observed with kinks that we initially have noticeably short bounces that become longer as the simulation evolves \cite{kinkWobble}.

Next, we study a phase space of solutions to help identify any patterns in the behaviour of the excited 2-vortex scattering. We find solutions for a range of initial phases and initial velocities and hence generate a phase space of solutions, detailing the number of bounces as the number of times the vortices scatter through each other. 
\begin{figure}
\centering
\begin{subfigure}[b]{\textwidth}
    \centering
    \includegraphics[width=\textwidth]{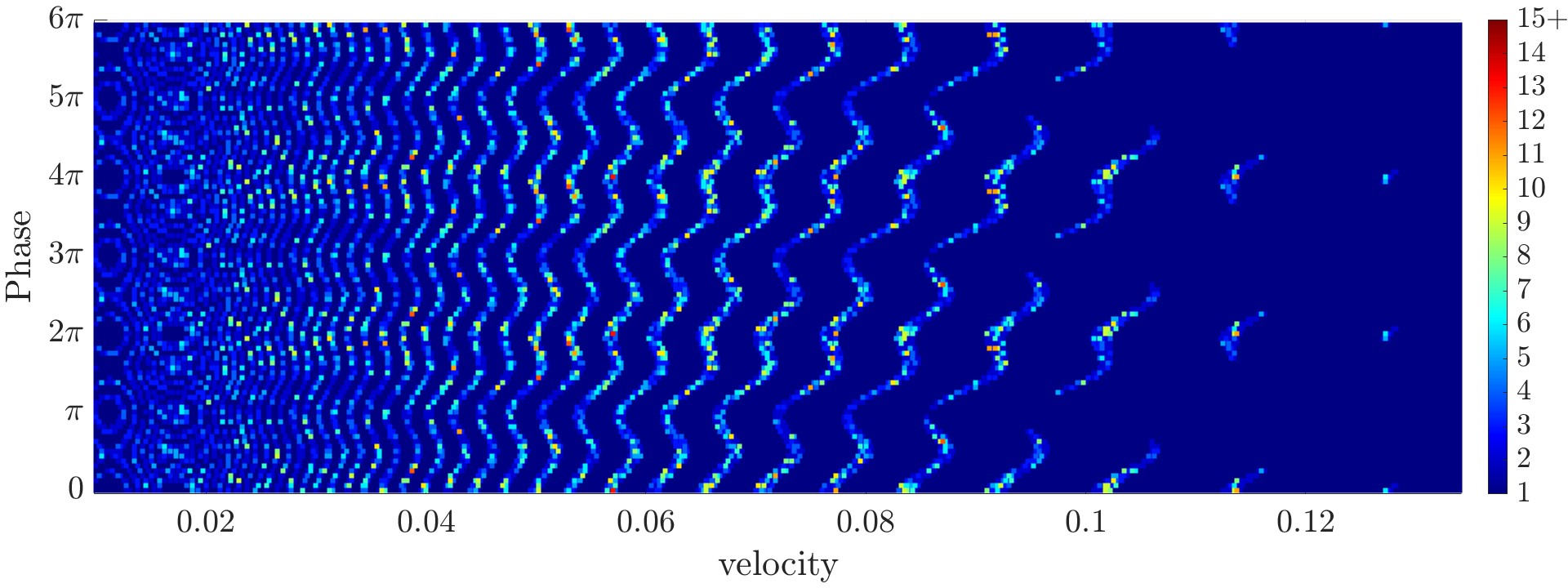}
\caption{Initial phase altered by the ansatz \eqref{Ansatz2}}
\label{fig:phaseSpaceNoNorm}
\end{subfigure}
\begin{subfigure}[b]{\textwidth}
    \centering
    \includegraphics[width=\textwidth]{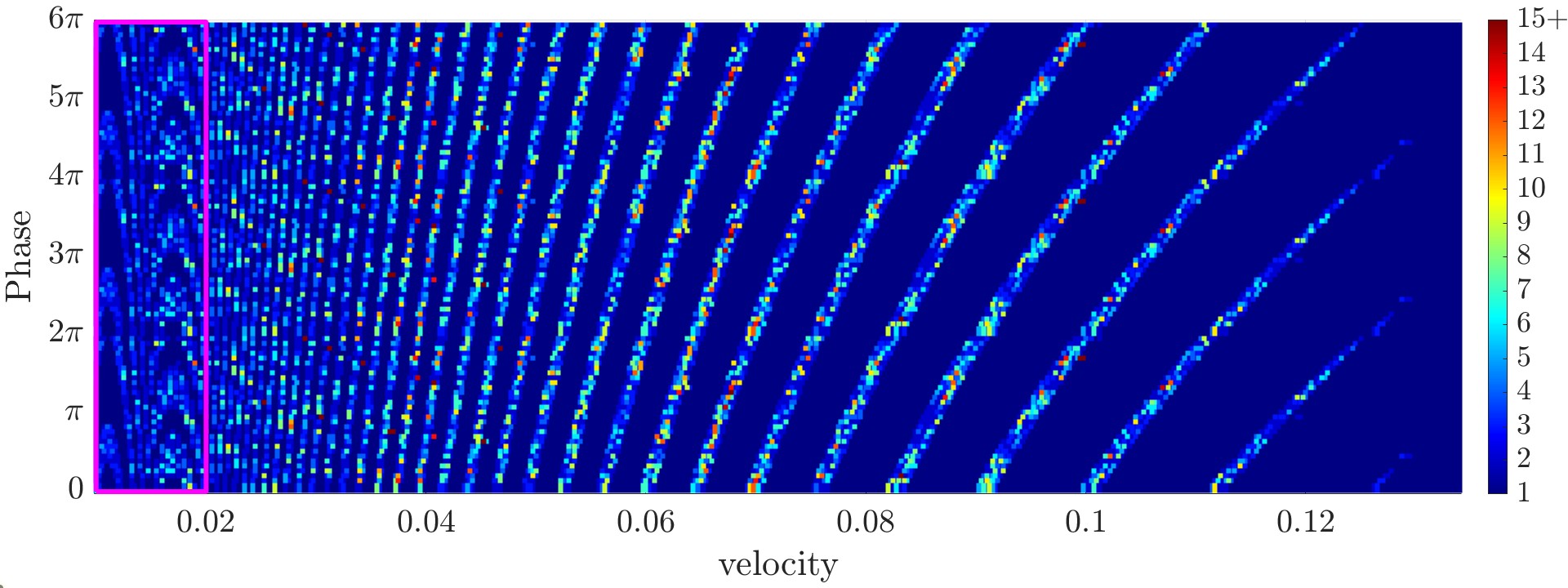}
\caption{Initial phase altered numerically using a displacement shift and evolving through time.}
\label{fig:phaseSpaceDispShift}
\end{subfigure}
\caption{Phase space of excited vortex scattering solutions. We show solutions for different initial velocity and initial phase for fixed $\epsilon = 0.9$. The dark blue space indicates solutions that only have one bounce, i.e. the vortices scatter only once, which is the normal behaviour for vortices at critical coupling. The number of bounces is represented as a heat plot for the colour of each simulation, shown by the colour bar. The data are plotted three times along the $y$ axis since the phase coordinate is cyclic, allowing us to get a clearer picture of the behaviour of the phase space.}
\label{fig:phaseSpaceEps}

\end{figure}

\figref{fig:phaseSpaceDispShift} shows a sample of solutions for a set of initial phases $\sigma(0) \in [0,2\pi)$, and initial velocities $v_{\rm in} \in [0.01,0.13]$. The number of bounces is indicated by the colour. The $y$-axis has been extended to be in the range $[0,6\pi)$. 

We observe in \figref{fig:phaseSpaceDispShift} regions of solutions that have multi-bounce scattering. We also observe in-between these regions sets of solutions that only scatter once. We find a fractal structure of multi-bounce solutions. We see that the lines of solutions that have multi-bounces also have a curvature, rather than a fixed slope. This is quite intuitive, as we have a series of lines of decreasing gradient; however, we can clearly see that the lines curve. This means that changing the initial phase of the mode is equivalent to changing the initial velocity, up to a critical value where the initial velocity dominates the interaction of the vortices, and they always escape. We can see that this critical velocity is around the region of $v_{\mathrm{in}} = 0.13$. However, this is only a rough approximation. Extending the phase space in the $y$-direction also allows us to see more easily that for any given initial velocity below the critical region, you can always choose an initial phase such that the vortices scatter more than once.

\begin{figure}
    \centering
    \includegraphics[width = 0.5\textwidth]{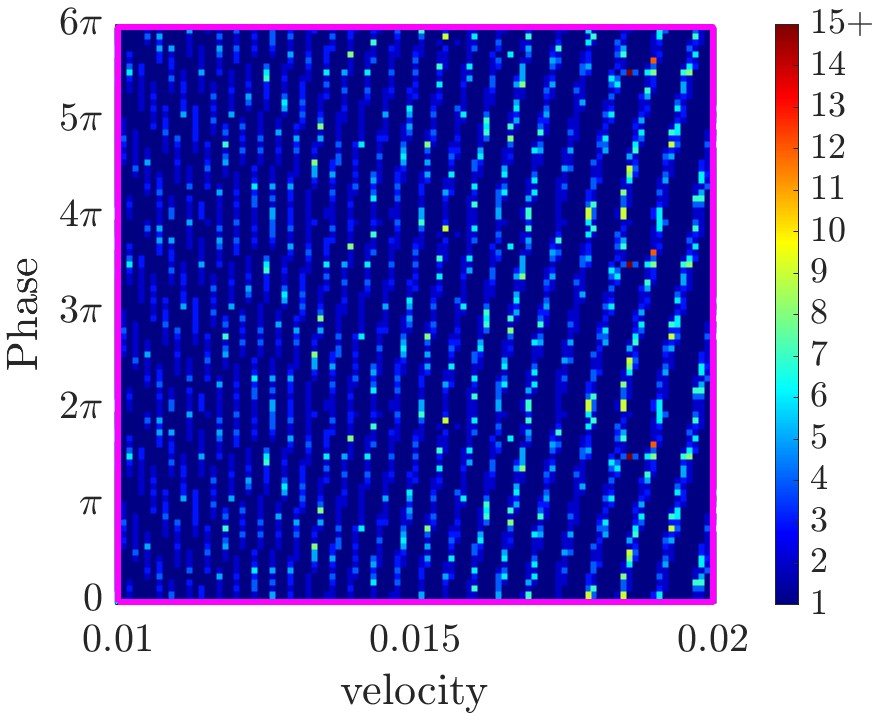}
    \caption{A higher resolution plot of the phase space plot. We plot the highlighted region of \figref{fig:phaseSpaceDispShift}, but using a smaller step in $v_{\rm in}$.
    }
    \label{fig:phaseSpace_zoom}
\end{figure}

For low velocities, the resolution of the phase space is too small to reveal the full structure of the phase space, hence the presence of the parabola in the parameter space of solutions \figref{fig:phaseSpaceDispShift} is a result of the resolution of the data. We can observe \figref{fig:phaseSpace_zoom} to show that the fractal structure of repeated lines is observed for small velocities, but due to the increasing slope of this pattern at low velocities, it is hard to capture the pattern at low velocities as the lines become more vertical and narrow, meaning that it is easy to miss when scanning the parameter space.

We can hence see that the whole phase space shows a fractal structure of regions with multiple bounces, ranging from 2 to 30 bounces. The number of bounces does not appear to have any correlation to the phase space at large, but it could be argued that the resolution of the diagram is too low to give a definitive answer. We now turn to the question of why this fractal pattern appears. We can surmise that this is a result of the phase of the shape mode altering the state of the interaction for different velocities. We see that periodically, there are these dark blue regions (solutions that only scatter once) and then thin slices of solutions with multiple bounces, increasing in width and decreasing in slope as the initial velocity increases. Furthermore, these factors appear to be constant for each region with respect to the phase. 

As stated above in Section \ref{sec:numerics}, we can alter the phase of the mode in two separate ways. We can see the phase space of solutions for both these methods in \cref{fig:phaseSpaceDispShift,fig:phaseSpaceNoNorm}. \figref{fig:phaseSpaceDispShift} shows solutions where the initial phase of the shape mode has been changed by shifting the initial vortex position $d_i$, and numerically evolve to alter the initial phase, and \figref{fig:phaseSpaceNoNorm} shows solutions where the initial phase is changed using the ansatz \eqref{Ansatz2}.

Due to the dependence of the energy on $\epsilon$ in the ansatz \eqref{Ansatz2}, which is maximal at a $\pi-$shift, we can see in the \figref{fig:phaseSpaceNoNorm} that the pattern of the results deviates most at this value, showing that the difference between these plots is an artifact of this phase dependence on the energy. We can hence assume that the plots should be identical, except that we have this deviation because the method used. 

We now discuss other initial amplitudes of the excitation. Take, for example, $\epsilon = 0.5$, i.e. $A(0) = 0.097.$  For this initial amplitude, the mode decays extremely slowly, hence non-linear effects are smaller. We find that, for small velocities, the vortices escape after one bounce. Hence, we can assume that for this amplitude, the scattering is dominated by the velocity, and the mode excitation causes little interaction between the vortices as they scatter. This gives further evidence to the proposition that the vortices escape the bound state due to the decay of the excitation, as if the amplitude is too small initially; they do not bounce more than once.
Therefore, we examine one more initial amplitude between these two values already discussed and take $\epsilon = 0.75$, such that $A(0) \approx 0.219$, which also decays slowly. We can see in \figref{fig:decayA} that this choice of $\epsilon$ corresponds to an initial amplitude of approximately $60\%$ of the previous amplitude discussed, where $\epsilon = 0.9$ and $A(0) = 0.317$. 
\begin{figure}
\centering
\includegraphics[width=\textwidth,height=16em]{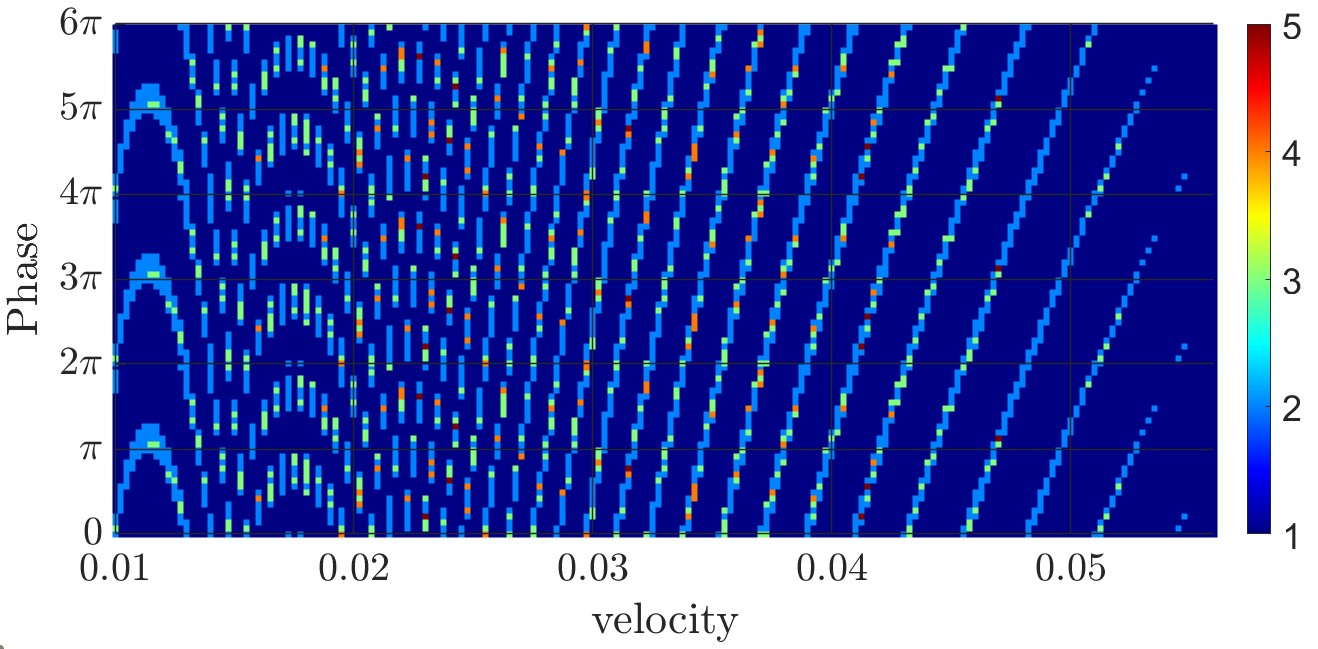}
\caption{Parameter Solution space detailing the space of solutions computed using a $2^\mathrm{nd}$ order Leapfrog method for time evolution. We show solutions for different initial velocity and initial phase for fixed $\epsilon = 0.75$. The dark blue space indicates solutions that only have one bounce, i.e. the vortices scatter only once, which is the normal behaviour for vortices at critical coupling. The number of bounces is represented as a heat plot for the colour of each simulation, shown by the colour bar. The data are plotted three times along the $y$ axis since the phase coordinate is cyclic, allowing us to get a clearer picture of the behaviour of the phase space.}
\label{fig:phaseSpaceEps0.3}
\end{figure}

We see in \figref{fig:phaseSpaceEps0.3} that we have the same fractal structure dominating the phase space. There are some key differences between the phase space of solutions with $\epsilon = 0.9$ and $\epsilon = 0.75$. Firstly, we observe in \figref{fig:phaseSpaceEps0.3} that there are only one bounce windows after an initial velocity of $v_{\mathrm{in}} \approx 0.055$. This suggests that the interaction imposed by the mode is weaker than the strength of the initial velocity, further supporting the conjecture that the mode requires a certain amount of energy to dominate the interaction. We further see that the fractal lines are narrower in \figref{fig:phaseSpaceEps0.3} than in \figref{fig:phaseSpaceDispShift}. However, they are significantly closer together, which could suggest that this set of solutions is just a scaled set of solutions compared to \figref{fig:phaseSpaceDispShift}. Note that, for small velocities, the line pattern is harder to see. This is due to the resolution of the phase space. With higher resolution, this part of the diagram would appear to fit the pattern of the rest of the data.

We now briefly explore the scattering of vortices where the excitations for each vortex are out of phase with each other. Consider the 2-vortex scattering with initial velocity $v_{\mathrm{in}} = 0.01$, and initial amplitude of the mode $A(0) = 0.097$. We have stated previously that we do not get any bounce windows when the vortices are in phase; however, we observe some interesting behaviour when we consider a relative phase of $\sigma = \pi$. 
\begin{figure}
\centering
\includegraphics[width = \textwidth]{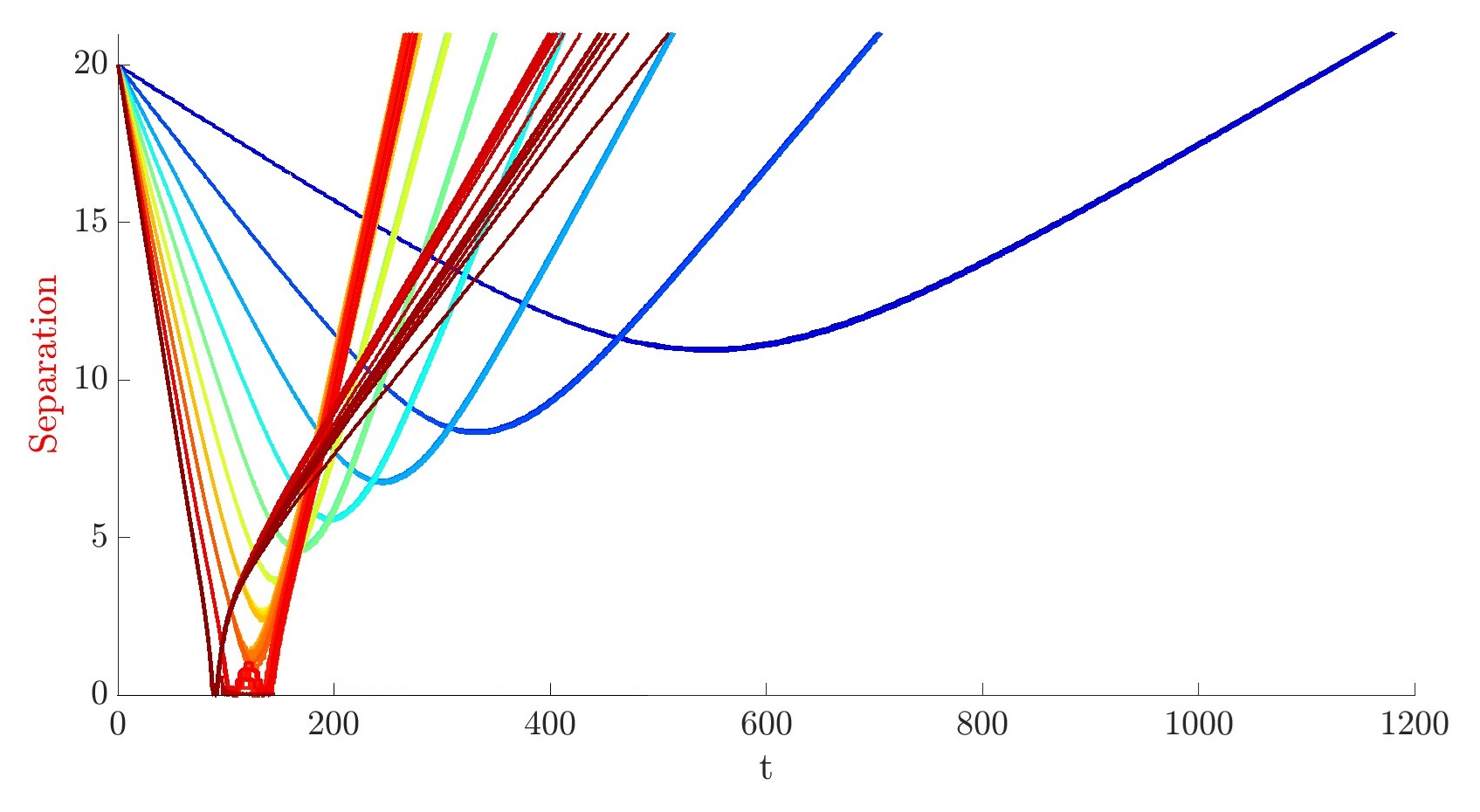}
\caption{Separation of the zeros of two vortices with initial velocity $v_{\mathrm{in}} = [0.01,0.1]$ and initial magnitude of the perturbation $\epsilon = 0.5$. The vortices are out of phase in terms of the shape mode by $\sigma = \pi$.}
\label{fig:0.5pi_oop}
\end{figure}

\figref{fig:0.5pi_oop} shows the separation of the zeros of the Higgs field for 2-vortex scattering with an excited shape mode. However with a relative phase between the shape modes per vortex of $\sigma = \pi$. The colours indicate the velocities. Dark blue corresponds to a velocity of $v_{\mathrm{in}} = 0.01$ and the dark red corresponds to an initial velocity of $v_{\mathrm{in}} = 0.1$, with the colours in between interpolating between these values with a step of $v_{step} = 0.01$. For small velocities we see a clear repulsion between the vortices, as they do not scatter. We can see from the slope of the separation that the vortices slow down when they become close, and then repel, as we can see the change in direction of the lines. We can see that the faster the velocity, the closer the vortices come together before repelling, suggesting that the repulsion is not dependant on the velocity, but fighting against it. This figure illustrates behaviours of type \rom{2} superconductors, which is where vortices are repulsive.
Although not presented here, we find that with a relative phase of $\sigma = \frac{\pi}{2}$ this mode is highly attractive, with solutions with up to 50 bounces. This is intriguing, as we previously only observed one-bounce windows with this initial amplitude. This implies that the behaviour of these vortices is highly dependent on the phase of the mode, since breaking the symmetry drastically alters the results, as seen in \figref{fig:0.5pi_oop}, where the vortices repel. More research will be conducted on this out-of-phase scattering to try and understand the effect of breaking the symmetry of the mode.

\section{\label{sec:conclusions} Conclusions and future work}
In this paper we have provided a detailed outline of how to develop the numerics for Abelian-Higgs vortices in $(2+1)$ dimensions. We further showed how to excite the vortex shape mode that causes fluctuations in the gauge-invariant quantities. We discussed extensive results that explore the scattering of the excited vortices. 

Interestingly, we found that that excitation of the shape mode leads to the scattering exhibiting behaviour of both type \rom{1} and type \rom{2} superconductors. When both vortices are in phase, the attractive properties dominate for most initial phases. We have provided numerical and analytical evidence of the nature of the attractive and repulsive properties of the mode. We then sampled a phase space of solutions and found a fractal structure dependent on the initial phase of the shape mode and initial velocity of the vortices. We found that the number of bounces a solution exhibits show signs of chaotic behaviour and depends sensitively on the initial setup of the scattering.

An important result of this paper is that geodesic flow is not a valid approximation for vortex scattering with excited shape modes, as it does not explain attraction due to excitation. During the production of this paper, models were developed in \cite{CCM} to explain the motion of vortices with excited shape mode, whereby geodesic flow on $\mathcal{M}$ is modified by a potential. Interestingly, a fractal pattern is also found in \cite{CCM}, whereby multiple bounces in 2-vortex collisions are observed, which is in complete agreement with the result presented in this paper.

This paper opens many avenues for future work. We are planning to study the dynamics of vortices of higher multiplicity, specifically, multi-vortex scattering at critical coupling. This has many possibilities because higher degree vortices also have more normal modes to excite. We also consider scattering excited vortices with a non-zero impact parameter. Initial calculations show that the vortices can orbit each-other. Remarkably, we can further study the same 2-vortex scattering with excited shape mode briefly discussed in this paper, namely where the vortices are $\pi$ out of phase. This is work in preparation \cite{SWPaper}, and we find spectral walls in the Abelian Higgs model.

\appendix
\section{\label{sec:appendixDerrick} Derrick's Scale Approximation}
This section seeks to show that the shape mode can be well approximated by a Derrick's scaling of the fields. First, we show the initial construction used to excite the mode this way. We apply Derrick's scaling argument and consider a spatial rescaling of the form,
\begin{equation}
x \mapsto \mu x = \tilde{x},
\label{eq:rescaling}
\end{equation}
where $\mu$ is the Derrick's scaling factor.
The Higgs field $\Phi$ scales as 
\begin{equation}
    \tilde{\Phi} = \Phi(\mu x).
\end{equation}
We are working in a gauge theory; hence we require that both terms of the covariant derivative $D_j\tilde{\Phi}$ scale consistently. Since \begin{equation}\partial_j\tilde{\Phi} = \mu(\partial_j\Phi)(\mu\ x),
\end{equation} we impose that the gauge potential $A_{\mu}$ scales as 
\begin{equation}
\tilde{A}_{\mu} = \mu A_{\mu}(\mu x).
\end{equation}
It is important to perform the Derrick's scaling on the fields before applying the Lorentz transformation, and the resulting initial condition is as follows,
\begin{equation}
	\begin{split}
	&\hat{\Phi}(t,x_1,x_2) = \left(\gamma(\mu x_1+vt)+i\mu x_2\right)^NF(\gamma^2(\mu x_1+vt)^2+\mu^2x_2^2),\\
	&\hat{A}_{mu}(t,x_1,x_2) = (\hat{A}_0, \hat{A}_1, \hat{A}_2) = \begin{pmatrix}Nv\gamma \mu^2 x_2G(\gamma^2(\mu x_1+vt)^2 + \mu^2x_2^2) \\ -N\gamma \mu^2 x_2G(\gamma^2(\mu x_1+vt)^2 + \mu^2x_2^2) \\ N\gamma\mu(\mu x_1+vt)G(\gamma^2(\mu x_1+vt)^2 + \mu^2x_2^2) \end{pmatrix},
	\end{split}
 \label{Initial_Condition}
\end{equation}
Hence we have an initial configuration for our two-dimensional dynamical numerics, detailing a axially symmetric vortex with an initial velocity and a Derrick mode excitation.

As stated above, we find the frequency of the shape mode to be $\omega^2_{\rm Lin} = 0.777476$. By studying the potential energy of the Derrick's scaled solution, we find a frequency of the approximated mode to be $\omega^2_{\rm Derrick} = 0.770076$, which is within $1\%$ of the frequency found through the linearisation of the full field theory. This gives us evidence that Derrick's scaling the solution is indeed a good approximation to the shape mode.

  We can calculate the 2-dimensional norm of the perturbation for both methods to model how well the Derrick's scaling approximates the mode.
\begin{equation}
    \label{eq:norm}
    \langle \mathbf{f},\mathbf{g} \rangle = \int \mathbf{f}\cdot \mathbf{g} \ d^2x,
\end{equation}
where $\mathbf{f}$ and $\mathbf{g}$ are vectors of the Higgs field and gauge fields for the Derrick's scale perturbation, and the linearisation perturbation respectively, such that
\begin{align}
    \mathbf{f} &= (\tilde{\psi}_1(x,y),\tilde{\psi}_2(x,y),\tilde{\chi}_1(x,y),\tilde{\chi}_2(x,y))^T,\\
    \mathbf{g} &= (\psi_1(x,y),\psi_2(x,y),\chi_1(x,y),\chi_2(x,y))^T.\\
\end{align}
The Derrick's scale mode approximation is a particularly good approximation to the linearisation, agreeing up to $95\%$ for small perturbations. As the perturbation grows larger, we see that the scale approximation begins to become a less accurate approximation. However, we still see for significantly high $\mu$, that the approximation covers $91\%$ of the linearisation, which is still considerably accurate. This gives us confidence that the scale approximation is a suitable method for exciting the mode. 

It is a useful result that the shape mode can be approximated by the Derrick's scaling. It has been shown in \cite{Hindmarsch} how to find eigenfunctions to excite the linear mode for all coupling $\lambda$. However, the method chosen in this paper follows procedures outlined in \cite{alonso2023spectral}, which outline a method for finding eigenfunctions for linear vortex modes. However only for critical coupling. The benefit of exciting the shape mode using a Derrick's scaling is not only a simpler procedure, but it is applicable to not only critical coupling, but also for all $\lambda$, hence this method could be applied for all solitons, including those where the linearisation is not yet known.

By approximating the shape mode by a Derrick's scaling, we also begin to gain an understanding of the properties of interaction of the mode. We can begin to study the long-range interaction of these excited vortices. 

Let \begin{equation}
    \Phi(x) = e^{i N \theta}f(\rho) = \frac{x_1+ix_2}{\rho^N}f(\rho),
\end{equation}
\begin{equation}
    \mathbf{A} = \begin{pmatrix}
        0 \\
        \frac{-x_2}{\rho^2}a_{\theta}(\rho) \\
        \frac{x_1}{\rho^2}a_{\theta}(\rho)
    \end{pmatrix},
\end{equation}
where $\rho$ is the radial coordinate such that $\rho = \sqrt{x_1^2+x_2^2}$, and $f(\rho)$ and $a_{\theta}(\rho)$ are profile functions, such that 
\begin{equation}
    f(\rho) = \rho^N F(\rho^2),
\end{equation}
\begin{equation}
    a_{\theta}(\rho) = \rho^2 G(\rho^2).
\end{equation}
We can linearise these profile functions near infinity, as shown in \cite{manton2004topological, Manton:E_int}, such that we get the following decaying solutions with asymptotic expressions for $|\rho| \gg 1$,
\begin{equation}
    f(\rho) \approx 1 - \frac{q}{2\pi}K_0(\sqrt{\lambda}\rho),
    \label{eq:besselPhi}
\end{equation}
\begin{equation}
    a_{\theta}(\rho) \approx N - \frac{m}{2\pi}\rho K_1(\rho),
    \label{eq:besselA}
\end{equation}
where $K_0(\rho)$ and $K_1(\rho)$ are modified Bessel functions, with leading exponential term $\sqrt{\frac{\pi}{2\rho}}e^{-\rho}$ for large $\rho$.

We find the coefficients $q$ and $m$ using numerical techniques. Furthermore, we can find these coefficients as the vortex wobbles, and track the changes in the coefficient, to show how the Derrick mode excitation provides an interactive force.

The interaction energy can be calculated as follows,
\begin{equation}
    E_{int}(s) = -\frac{q^2}{2\pi}K_0(\sqrt{\lambda}s) + \frac{m^2}{2\pi}K_0(s).
\end{equation}

We can interpret the interaction energy as follows. The $q$ term in the interaction energy is negative, hence the length scale associated with the scalar field produces an attraction. Additionally, the $m$ term is positive; hence the length scale associated with the magnetic field produces a repulsion. Thus, when the interaction energy is positive, the magnetic field dominates the interaction, and the vortices exhibit a repulsive force, whereas when the interaction energy is negative, the scalar field dominates the interaction, and the vortices exhibit an attractive force.

If we now consider the spatial rescaling \eqref{eq:rescaling}, then for $\rho \gg 1$, the equations \eqref{eq:besselPhi},\eqref{eq:besselA} rescale as,
\begin{equation}
    f(\rho) \to \tilde{f}(\rho) = f(\mu\rho) \approx 1 - \frac{q}{2\pi}K_0(\mu\rho),
    \label{eq:rescaleBessel1}
\end{equation}
\begin{equation}
    a_{\theta}(\rho) \to \tilde{a_{\theta}}(\rho) = a_{\theta}(\mu\rho) \approx 1 - \frac{m}{2\pi}\tilde{\rho}K_1(\mu\rho).
    \label{eq:rescaleBessel2}
\end{equation}
If we consider the magnetic field, we have that 
\begin{equation}
    \begin{split}
        B &= \frac{1}{\rho}\frac{\partial a_{\theta}(\rho)}{\partial \rho}\\
        \tilde{B} &= \frac{1}{\tilde{\rho}}\mu\frac{\partial \tilde{a_{\theta}}(\rho)}{\partial \rho}\\
        &=-\frac{\mu}{\tilde{\rho}}\frac{m}{2\pi}(\mu K_1(\mu\rho)+\mu\tilde{\rho}K_1'(\mu\rho))\\
        &=\mu^2\frac{m}{2\pi}(\frac{1}{\tilde{\rho}}K_0'(\mu\rho)+K_0''(\mu\rho))\\
        &=\mu^2\frac{m}{2\pi}K_0(\mu\rho).
    \end{split}
\end{equation}
Hence, the interaction energy at critical coupling becomes
\begin{equation}
    E_{int}(R) = -\frac{q^2}{2\pi}K_0(\mu R) + \mu^2\frac{m^2}{2\pi}K_0(\mu R),
\end{equation}
where $R$ is the separation between two vortices.

Therefore, we see that when $\mu < 1 $ the magnetic interaction is weaker, and hence there will be an attraction. Moreover, when $\mu > 1$, the magnetic interaction is stronger, and hence there will be a repulsion between the vortices.
It is important to note that we have also developed the numerics to excite the shape mode by a Derrick's scaling of the fields. Indeed, using a Derrick's scaling to excite the mode is only an approximation, and hence there is more radiation in the system when the excitation is carried out this way. However, it is much easier numerically to include a mode excitation of this form.

We can compare this with the numerically calculated interaction energy, such that $E_{int} = V_2(\Phi,A_{\mu}) - 2V_1(\Phi,A_{\mu})$, where $V_2$ is the potential energy of a two-vortex system, where the vortices have been pinned at fixed positions of $d_i = \pm 8$, and $V_1$ is the potential energy of a single vortex. We have calculated this interaction energy for the shape mode discussed but have also calculated the interaction energy for the same mode excited instead with a Derrick's scaling, with approximately the same initial amplitude of the mode.
\begin{figure}
    \centering
    \includegraphics[width=0.75\textwidth]{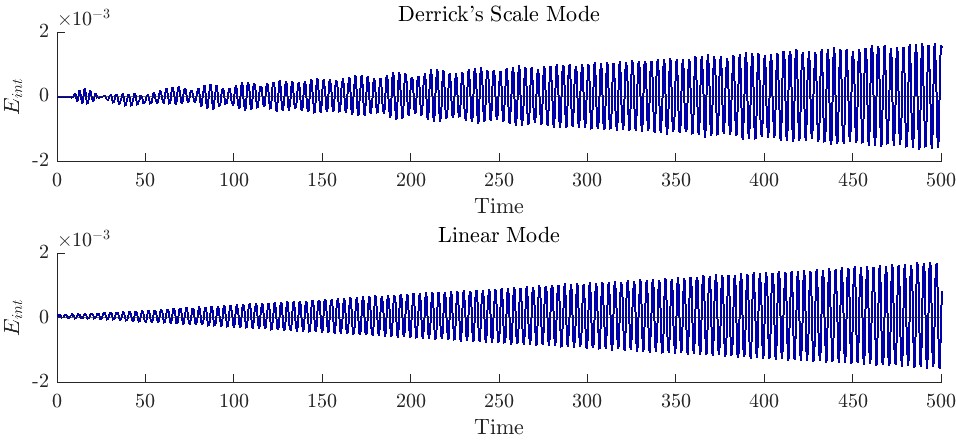}
    \caption{Plot to show the interaction energy for a 2-vortex system pinned at $d_i = \pm8$, with initial amplitude of the mode $A = 0.006$, corresponding to $\epsilon = 0.125$ and Derrick's scale factor $\mu = 0.93$.}
    \label{fig:InteractionEnergy}
\end{figure}

We can see in \figref{fig:InteractionEnergy} that the fluctuations in the interaction energy are more stable where the mode has been excited using the linearisation techniques; however, the interaction energy calculated from the Derrick mode excitation is still a really good approximation to that of the linearisation, showing further that this mode can be accurately modelled by a Derrick scaling.

We show the phase space of solutions for a mode excitation of this form
\begin{figure}
\centering
\includegraphics[width=\textwidth]{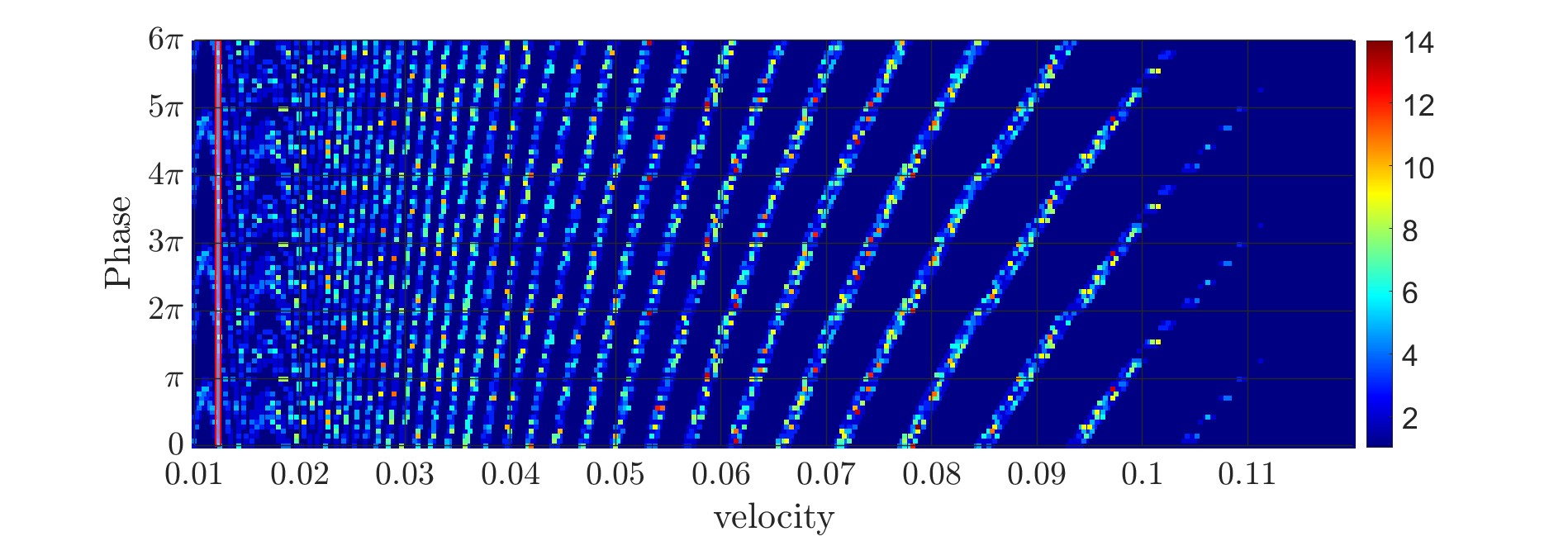}
\caption{Phase space of scattering solutions. We show solutions for different initial velocity and initial phase for fixed Derrick factor $\mu = 0.7$. The dark blue space indicates solutions that only have one bounce, i.e. the vortices scatter only once, which is the normal behaviour for vortices at critical coupling. The number of bounces is represented as a heat plot for the colour of each simulation, shown by the colour bar. The data are plotted three times along the $y$ axis since the phase coordinate is cyclic, allowing us to get a clearer picture of the behaviour of the phase space.}
\label{fig:phaseSpace}
\end{figure}
\figref{fig:phaseSpace} shows that we can observe the same behaviour as in \figref{fig:phaseSpaceEps}, showing that using a Derrick's scaling to excite the mode is a good enough approximation.

\section{\label{sec:appendixBC} Boundary Conditions}
\label{boundaryconditions}
We impose natural boundary conditions \cite{speight2019chiral}, so that radiation may leave the system by passing through the boundary. We denote the dynamical fields collectively as $\xi_a$, $a = 0,..,5$, consisting of the real and imaginary components of $\Phi$, as well as the 3 components of the vector gauge potential. We take the variation of the energy functional with respect to $\xi_a$, so that the energy varies as:
\begin{equation}
	\delta E = \int_{\Omega}\left(\frac{\partial E}{\partial \xi_a} -\partial_i\left(\frac{\partial E}{\partial(\partial_i\xi_a)}\right)\right)\delta\xi_a + \int_{\partial\Omega}\left(- n_i\frac{\partial E}{\partial(\partial_i\xi_a)} \right)\delta\xi_a
\end{equation}
where $\Omega$ is the finite domain that we perform our simulations on, and $\partial\Omega$ is the boundary of the domain. Furthermore, the divergence theorem has been used such that the flux of the variation of $E$ through the boundary curve $\partial\Omega$ is the same as the surface integral of the divergence of the variation of $E$ across the entire region $\Omega$. It should be noted that $\mathbf{n}$ is the inward pointing normal to $\partial\Omega$. We require that $\delta E = 0$ be such that $\xi_a$ satisfies the Euler-Lagrange equations in $\Omega$. Henceforth, we have the boundary conditions.
\begin{equation}
\label{eq:BC}
	n_i\frac{\partial E}{\partial(\partial_i\xi_a)} = 0
\end{equation}
on the boundary $\partial\Omega$.
First, we consider the boundary $x_1=\pm\infty$. For the energy \eqref{eq:staticEnergy}, the boundary condition \eqref{eq:BC} reduces to
\begin{equation}
\begin{split}
	\begin{pmatrix} 1 \\ 0 \end{pmatrix} \cdot \begin{pmatrix}
	\partial_1\phi_1 + A_1\phi_2 \\
	\partial_2\phi_1 + A_2\phi_2
	\end{pmatrix} &= 0 \Rightarrow \partial_1\phi_1 = -A_1\phi_2,\\
	\begin{pmatrix} 1 \\ 0 \end{pmatrix} \cdot \begin{pmatrix}
	\partial_1\phi_2 - A_1\phi_1 \\
	\partial_2\phi_2 - A_2\phi_1
	\end{pmatrix} &= 0 \Rightarrow \partial_1\phi_2 = A_1\phi_1,\\
	\begin{pmatrix} 1 \\ 0 \end{pmatrix} \cdot \begin{pmatrix}
	\partial_1A_2 - \partial_2A_1 \\ 0
	\end{pmatrix} &= 0 \Rightarrow \partial_1A_2 = \partial_yA_1\\
	\begin{pmatrix} 1 \\ 0 \end{pmatrix} \cdot \begin{pmatrix}
	\partial_0A_1 - \partial_1A_0 \\ 0
	\end{pmatrix} &= 0 \Rightarrow \partial_1A_0 = \partial_0A_1.
\end{split}
\end{equation} 
We must also consider the boundary $x_2=\pm\infty$
\begin{equation}
\begin{split}
	\begin{pmatrix} 0 \\ 1 \end{pmatrix} \cdot \begin{pmatrix}
	\partial_1\phi_1 + A_1\phi_2 \\
	\partial_2\phi_1 + A_2\phi_2
	\end{pmatrix} &= 0 \Rightarrow \partial_2\phi_1 = -A_2\phi_2,\\
	\begin{pmatrix} 0 \\ 1 \end{pmatrix} \cdot \begin{pmatrix}
	\partial_1\phi_2 - A_1\phi_1 \\
	\partial_2\phi_2 - A_2\phi_1
	\end{pmatrix} &= 0 \Rightarrow \partial_2\phi_2 = A_2\phi_1,\\
	\begin{pmatrix} 0 \\ 1 \end{pmatrix} \cdot \begin{pmatrix}
	0\\ \partial_1A_2 - \partial_2A_1
	\end{pmatrix} &= 0 \Rightarrow \partial_1A_2 = \partial_2A_1\\
	\begin{pmatrix} 0 \\ 1 \end{pmatrix} \cdot \begin{pmatrix}
	\partial_0A_2 - \partial_2A_0 \\ 0
	\end{pmatrix} &= 0 \Rightarrow \partial_2A_0 = \partial_0A_2.
\end{split}
\end{equation} 
Furthermore, we are working in a discretised version of a continuous theory, so we must also discretise our boundary conditions, which give us equations for ghost points, which are temporary points that exist past the boundary. These allow us to calculate the $1^{\rm st}$ and $2^{\rm nd}$ degree finite difference derivatives on the boundary. 

These boundary conditions can be summarised such that the covariate derivative tends to zero normal to the boundary at infinity, as well as the magnetic field, i.e.
\begin{equation}
	\begin{split}
		\mathbf{n}\cdot(\nabla - i\mathbf{A})\Phi = n_i\cdot D_i\Phi &= 0\  \mathrm{on}\ \partial\Omega\\
		\mathrm{curl}\mathbf{A} = \nabla \times \mathbf{A} = B &= 0 \ \mathrm{on}\ \partial\Omega\\
	\end{split}	
\end{equation}
where the gauge potential $\mathbf{A}$ is a 4 component 1-form, with the $z-$dependence set to zero.

We must impose a further constraint on the boundary such that the $1^{\rm st}$-order time derivative of the electric potential $A_0$ goes to $0$ on $\partial\Omega$, i.e. $\partial_0A_0 = 0$. This constraint is necessary for numerical stability.

\section*{\label{sec:Acknowledgements} Acknowledgements}
The authors thank Jennifer Ashcroft for interesting discussions and her thesis titled ``Topological solitons and their dynamics'' \cite{ashcroft2017topological}.
All numerical calculations were run using the high performance computing systems provided by the University of Kent. Morgan Rees acknowledges the UK Engineering and Physical Sciences Research Council (EPSRC) for a PhD studentship. Thomas Winyard would like to thank the School of Mathematics at the University of Edinburgh for funding his postdoctoral position.

\bibliographystyle{plain}
\bibliography{ExcitedVortexScattering}
\end{document}